\documentclass[journal]{IEEEtran}

\usepackage[pdftex]{graphicx}

\usepackage{array}
\usepackage{multirow}
\usepackage{bm}
\usepackage{mdwmath}
\usepackage{xcolor,soul,framed}
\usepackage{cite}
\usepackage{eqparbox}
\usepackage{url}
\usepackage{amsmath}
\usepackage{amssymb}
\usepackage{textcomp}
\usepackage{tikz}
\usepackage{bm}
\usepackage{dblfloatfix}

\newcommand{\branches}{$\mathcal{E}$}
\newcommand{\branchesm}{\mathcal{E}}

\newcommand{\buses}{$\mathcal{N}$}
\newcommand{\busesm}{\mathcal{N}}

\newcommand{\meas}{$\mathcal{M}$}
\newcommand{\measm}{\mathcal{M}}

\newcommand{\timeseries}{$\mathcal{T}$}
\newcommand{\timeseriesm}{\mathcal{T}}

\begin{document}

\bstctlcite{IEEEexample:BSTcontrol}
\title{Phase Identification of Distribution System Users Through a MILP Extension of State Estimation}

\author{Marta~Vanin,~\IEEEmembership{Member,~IEEE, }
 Tom~Van~Acker,	Reinhilde~D'hulst, 
       and~Dirk~Van~Hertem,~\IEEEmembership{Senior Member,~IEEE}

\thanks{M. Vanin, T. Van Acker, and D. Van Hertem are with the Research Group ELECTA, Department of Electrical Engineering, KU Leuven, 3001 Heverlee, Belgium. 
R. D'hulst is with VITO, Boeretang 200, 3400 Mol, Belgium.
M. Vanin, R. D'hulst, and D. Van Hertem are also with EnergyVille, Thor Park 8310, 3600 Genk, Belgium.
 Corresponding author: marta.vanin@kuleuven.be
 }
}
\maketitle

\begin{abstract}
To address the challenges and exploit the opportunities that the decarbonization of the energy sector is bringing about, advanced distribution network management and operation strategies are being developed. Many of these require accurate network models to work effectively, including user phase connectivity. However, such information is typically unavailable. This paper proposes a novel method to identify the phase connectivity of single- and three-phase distribution consumers using smart meter measurements. The method is based on state estimation and mixed-integer linear programming, and requires shorter measurement collection times compared to statistical and machine learning-based techniques, for the same level of accuracy. Such reduced measurement requirements accelerate the phase identification process, and the consequent creation of digital models for advanced network applications. If an ad-hoc measurement campaign is required for phase identification, its duration can be shortened and its costs reduced. If smart meters are installed, their measurement memory buffer may be sufficient for phase identification, without the need for additional data collection and management infrastructure. 
Extensive computational results are presented for publicly available low voltage feeders.
\end{abstract}

\begin{IEEEkeywords}
Distribution network, mixed-integer programming, parameter estimation, phase identification, state estimation
\end{IEEEkeywords}

\section{Introduction}\label{sec:introduction}
\subsection{Background and Motivation}

The increasing amount of distributed generation and the electrification of transport and heating systems are changing the nature of the distribution networks (DNs). 
These technologies can lead to technical issues, e.g., higher unbalance levels, but also have the potential to, e.g., provide flexibility for a more active network management. To this respect, extensive research has been conducted on, e.g., reactive power compensation~\cite{Liu2019}, PV curtailment~\cite{Nguyen2019}, phase reconfiguration~\cite{LiuReconfiguration}, and tap changer coordination~\cite{Gutierrez-Lagos}. The methods used in references~\cite{Liu2019}~-~\cite{Gutierrez-Lagos}, as any other active distribution network method that relies on power flow equations, requires accurate network data, including consumer phase connectivity. However, the latter is typically unknown~\cite{Primadianto}. As such, phase identification (PI) is crucial in the transition towards active distribution networks, and it has been largely addressed. A literature review of existing methods is performed hereafter. The causes behind the lack of connectivity information in today's distribution grid data sets, on the other hand, have been described in~\cite{GethCIRED23}. 

\subsection{Related Work}
Visual or manual verification of user connectivity is not feasible, given the huge geographical extension of DNs, and the fact that cables are often underground and/or not color-coded. Devices exist that can inject distinctive signals and detect on which phase they were generated~\cite{CairdPatent}, but these are excessively expensive to be deployed on a large scale system. Phasor measurement units (PMUs) have also been used to perform PI~\cite{WenmPMUs, Bariya}, but similar cost considerations apply. Smart meters (SMs), on the other hand, are more economical and are already being rolled out in DNs. Their use for PI has been extensively investigated and, unless specified otherwise, the references in the remainder of this section rely on SM measurements. One of the earliest works addressing SM-based PI~\cite{AryaMIP} consists of minimizing the difference between the time series of the measured transformer power supply and the total feeder demand per phase. To do so, three binary variables are assigned to every user (one per phase, indicating whether the user is connected to that phase or not), resulting in a mixed-integer programming (MIP) problem. Network losses are not modelled in~\cite{AryaMIP}, so the method is only effective when these are very limited. Most subsequent PI literature seems to refrain from MIP, possibly due to scalability issues. In~\cite{Pappu2018}, principal component analysis and graph theory are used to derive the network topology from energy measurements. Ref.~\cite{Xu2018} and~\cite{Tang2018} use power measurements from users and transformer, to which they apply spectral analysis and least absolute shrinkage and selection operator, respectively. Ref.~\cite{Jayadev} uses principal component analysis and conservation of energy constraints. In~\cite{GonzalezCagigal}, a method is proposed that relies on hourly energy data and nonlinear Kalman filters. Relying on power/energy measurements like~\cite{AryaMIP, Pappu2018, Xu2018, Tang2018, Jayadev, GonzalezCagigal} is convenient, as these are more often available than voltage measurements because they are collected by utilities for billing purposes. However, these methods can lead to inaccurate PI depending on the magnitude and pattern similarities of user demand, particularly in networks with many users. Moreover, multiple solutions may exist for methods that rely on the conservation of energy, like~\cite{AryaMIP, Jayadev}.

Using voltage time series is a popular alternative. Multiple researchers use them to classify users with k-means clustering-based methods~\cite{Arya2013, Wang2016, Simonovska, Short2013, Olivier}. Alternatively, the voltage time series correlation between the users and a reference, either the transformer~\cite{PezeshkiAUPEC} or a three-phase connection~\cite{PezeshkiISGT}, can be calculated. Each user is subsequently assigned to the reference's phase to which it correlates the most. Ref.~\cite{Foggo2020} proposes a supervised machine learning approach, which however requires an existing training data set. In general, voltage-based methods are less accurate for short and balanced feeders~\cite{Short2013}. Thorough comparisons of different statistical and clustering methods for PI can be found in~\cite{Therrien, Hoogsteyn2022}.

PI can be performed stand-alone, or as part of network topology identification for unbalanced networks. This consists of the data-driven reconstruction of the full network ``tree" e.g., using correlation methods~\cite{Liao2019}, or graphical models and conditional independence~\cite{Deka2020}. In this context, Bariya et al.~\cite{Bariya} propose a greedy algorithm which provides optimality guarantees and solves in polynomial time. However, it requires radiality, diagonal dominance of the impedance matrix and strong statistical assumptions on the correlation of nodal current injections. 

Note that the term ``topology identification" is also used to refer to the identification of switch states (open or close). These can change relatively often (e.g., during operations or maintanance actions) without being reported in real-time, and can be part of state estimation workflows, like in~\cite{Fernandes}. However, this is another problem and does not include PI.

Recently, a few papers re-introduced mixed-integer linear programming (MILP) for PI. Wang et al.~\cite{Wang2020} propose a maximum-likelihood-estimation approach with binary variables, but these are successively relaxed. Heidari et al.~\cite{Heidari-Akhijahani2021} propose a MILP technique with linearized power flow equations, minimizing the least absolute difference between measured values and system variables. To solve it for larger networks, an accelerated Benders decomposition is proposed. Contrary to most (if not all) statistical methods, MILP techniques are accurate in the presence of power injection from PV, which alters voltage patterns making correlation harder~\cite{Heidari-Akhijahani2021}. This seems an important advantage as the number of PV installations is increasing. However, we believe that the modelling, scalability and accuracy of~\cite{Heidari-Akhijahani2021} can be improved.

\subsection{Contributions and Comparison with Other Methods}

This work presents a novel method to perform PI in DNs, that requires SM measurements at end-users only. The method relies on state estimation (SE) and MILP. The LinDist3Flow (LD3F)~\cite{Sankur} linearization of unbalanced power flow equations is adopted. The method can be applied to both meshed and radial networks.

To the authors' knowledge, the first paper that includes full power flow equations for MILP-PI is~\cite{Heidari-Akhijahani2019} (and its follow-up~\cite{Heidari-Akhijahani2021}), and it is the state-of-the-art method which is most similar to the one proposed in this paper. The use of power flow equations for PI exposes it to errors in the cable impedance data, which may negatively affect the PI accuracy. Nevertheless, numerical results show that the proposed method is robust against this type of errors.  

The approach presented in our paper is more general than~\cite{Heidari-Akhijahani2019, Heidari-Akhijahani2021}, as it is based on proper SE rather than on the difference between variables and measurements. This allows to include different measurement types/measured quantities and measurement weights. In general, relying on SE is beneficial to filter measurement noise. Moreover,~\cite{Heidari-Akhijahani2021} relies on the linearization from~\cite{gharebaghi2019}, which assumes near-unitary voltage magnitudes, coefficient-based relations between power and voltage, and results in non-convexities when applied to PI, requiring additional approximations. The LD3F, on the other hand, accurately approximates low voltage (LV) system behaviour for other MILP problems, even in high-load conditions and sizeable networks~\cite{Vanin2020}, and does not result in non-convexities for PI. Furthermore, a formulation tightening technique is presented in Section~\ref{sec:tightening}, \textit{which is also novel}.

The main contributions of this work are the following:
\begin{itemize}
    \item A novel and effective MILP-SE-PI method is proposed. It is shown that larger networks can be decomposed for scalability purposes without compromising PI accuracy. 
    \item Alternative approaches for PI of three-phase users are presented and extensively compared. One of such approaches is novel, as well as a thorough discussion on PI for three-phase users. 
\end{itemize}

The existing literature focuses on single-phase user connectivity only. Of course, three-phase users are connected to all three phase conductors, but utilities usually ignore to which phase a certain SM channel measurement corresponds. Three-phase SM need to be ``aligned" to avoid errors in measurement-based calculations, e.g., SE, so the PI problem is extended for this purpose. Note that utilities do know which users are single- and which are three-phase~\cite{GethCIRED23}.


The proposed method presents a number of advantages, even when line impedance errors are present, that stem from the use of network equations. The main practical one is that satisfactory PI accuracy is achieved with shorter measurement time series in comparison to statistical methods. In addition to accelerating the system information acquisition process, this has the potential to reduce measurement campaign costs for the system operator. The MILP-SE-PI works with a combination of power and voltage measurements, which can compensate the drawbacks of voltage- or power-exclusive techniques. For instance, users that have identical power consumption patterns\footnote{This is an extreme case, unlikely to consistently occur throughout a time series.} can still be distinguished by their different voltage patterns, instead of generating two identical solutions like they would in~\cite{AryaMIP}. Furthermore, the global optimum of the MILP problem can be found\footnote{Note that this alone may not be sufficient to guarantee 100\% accurate PI (among other, measurement quantity and noise affect accuracy), but is at least physically explainable.}, whereas scalable k-means algorithm implementations are usually based on heuristics and can converge to local optima, making it harder to give a physical interpretation to the results~\cite{Bariya}. Finally, the use of SE naturally enables to filter measurement noise, reducing the impact of measurement errors on the PI.

The rest of the paper is organized as follows: Section~\ref{sec:mathematical_model} describes the mathematical model and implementation of the proposed MILP-SE-PI, and Section~\ref{sec:improvements} reports a series of techniques that can be used to obtain faster solutions. Section~\ref{sec:case_studies} describes the case studies and shows numerical results. Finally, Section~\ref{sec:conclusions} reports the main conclusions.






%
\section{Mathematical Model}\label{sec:mathematical_model}
\newcommand{\z}{$\mathbf{z}$}
\newcommand{\h}{$\mathbf{h}$}
\newcommand{\hhm}{$\mathbf{h}_m$}
\newcommand{\errv}{$\boldsymbol{\eta}$}
\newcommand{\vm}{$|U|$}
\newcommand{\va}{$\angle U$}
\newcommand{\vmjp}{$|U_{j,p}|$}
\newcommand{\vajp}{$\angle U_{j,p}$}
\newcommand{\vi}{$U^{\text{im}}$}
\newcommand{\vr}{$U^{\text{re}}$}
\newcommand{\ca}{$\angle I $}
\newcommand{\cax}{$\angle I_c $}
\newcommand{\cix}{$I^{\text{im}}_c$}
\newcommand{\crx}{$I^{\text{re}}_c$}
\newcommand{\cmx}{$|I_c|$}
\newcommand{\px}{$P_c$}
\newcommand{\qx}{$Q_c$}
\newcommand{\w}{$W$}

\newcommand{\cixm}{I^{\text{im}}_c}
\newcommand{\crxm}{I^{\text{re}}_c}
\newcommand{\cmxm}{|I|_c}
\newcommand{\vim}{U^{\text{im}}}
\newcommand{\vrm}{U^{\text{re}}}
\newcommand{\cam}{\angle I}
\newcommand{\caxm}{\angle I_c}
\newcommand{\vmm}{|U|}
\newcommand{\vmmpfi}{|U_i^{\text{pf}}|}
\newcommand{\vmmsei}{|U_i^{\text{se}}|}
\newcommand{\vam}{\angle U}
\newcommand{\pxm}{P_c}
\newcommand{\qxm}{Q_c}
\newcommand{\wm}{W}

\newcommand{\nodevar}{$\breve{x}$}
\newcommand{\edgevar}{$\tilde{x}$}
\newcommand{\nodevarm}{\breve{x}}
\newcommand{\edgevarm}{\tilde{x}}

\newcommand{\nodemeas}{$\breve{\mathcal{M}}$}
\newcommand{\edgemeas}{$\tilde{\mathcal{M}}$}
\newcommand{\nodemeasm}{\breve{\mathcal{M}}}
\newcommand{\edgemeasm}{\tilde{\mathcal{M}}}

\newcommand{\branchset}{$\mathcal{E}$}
\newcommand{\revbranchset}{$\mathcal{E}^{\text{R}}$}
\newcommand{\branchsetm}{\mathcal{E}}
\newcommand{\revbranchsetm}{\mathcal{E}^{\text{R}}}
\newcommand{\branchseti}{$\mathcal{E}_i$}
\newcommand{\revbranchseti}{$\mathcal{E}_i^{\text{R}}$}
\newcommand{\branchsetmi}{\mathcal{E}_i}
\newcommand{\revbranchsetmi}{\mathcal{E}_i^{\text{R}}}

\newcommand{\loadset}{$\mathcal{L}$}
\newcommand{\genset}{$\mathcal{G}$}
\newcommand{\loadseti}{$\mathcal{L}_i$}
\newcommand{\genseti}{$\mathcal{G}_i$}
\newcommand{\loadsetm}{\mathcal{L}}
\newcommand{\gensetm}{\mathcal{G}}
\newcommand{\loadsetmi}{\mathcal{L}_i}
\newcommand{\gensetmi}{\mathcal{G}_i}

\newcommand{\singlephaseusersetm}{\mathcal{U}^{1p}}
\newcommand{\threephaseusersetm}{\mathcal{U}^{3p}}

\newcommand{\N}{\textbf{\textcolor{black}{N}}}

\newcommand{\varspace}{$\mathcal{X}$}
\newcommand{\varspacem}{\mathcal{X}}

\newcommand{\wls}{WLS}
\newcommand{\rwls}{rWLS}
\newcommand{\wlav}{WLAV}
\newcommand{\rwlav}{rWLAV}

\newcommand{\Status}{s}
\newcommand{\Activation}{y}
\newcommand{\Deactivation}{z}
\newcommand{\ApparentPower}{S}
\newcommand{\ApparentPowerVector}{\mathbf{\ApparentPower}}
\newcommand{\ActivePower}{P}
\newcommand{\ActivePowerVector}{\mathbf{\ActivePower}}
\newcommand{\ForecastedActivePower}{\ActivePower^{\text{fx}}}
\newcommand{\GuaranteedActivePower}{\ActivePower^{\text{gtd}}}
\newcommand{\ReactivePower}{Q}
\newcommand{\ReactivePowerVector}{\mathbf{\ReactivePower}}
\newcommand{\ForecastedReactivePower}{\ReactivePower^{\text{fx}}}
\newcommand{\GuaranteedReactivePower}{\ReactivePower^{\text{gtd}}}

\IEEEeqnarraydefcolsep{0}{\leftmargini}


Let \buses \ and \branches \ be the set of a network's nodes (buses) and edges (branches), respectively. As any optimization problem, the proposed MILP-SE-PI is made of a set of variables and constraints, and an objective function. Let $\mathbf{x}$ and $\mathbf{y}$ be continuous variables. $\mathbf{x}$ contains the power flow variables that belong to the LD3F formulation's variable space, i.e., active and reactive power injections ($\mathbf{p}_i, \mathbf{q}_i \; \forall i \in \busesm$) and flows ($\mathbf{p}_{ij}, \mathbf{q}_{ij} \; \forall ij \in \branchesm$) and the squared voltage magnitude $\bm{ \omega }_i = \mathbf{|u|}_i^2 \; \forall i \in \busesm$. $\mathbf{y}$ are auxiliary variables required for the PI, as described later. Finally, $\bm{\delta}$ are the problem's integer variables, which are all binary. Bold mathematical symbols and letters represent vectors, if lowercase, and matrices, if uppercase. 

Let \meas \ be the set of available measurements, and let \timeseries \ be the timesteps for which all measurements are collected, i.e., $\timeseriesm = \{1, 2,..., T\}$, where, e.g., for an hour of quarter-hour measurements, $T$ = 4. The MILP-SE-PI problem can be summarized as:

\begin{IEEEeqnarray}{ l C l }
    \text{minimize} \; \;       &~& \sum_{\substack{t \in \timeseriesm }}\sum_{\substack{m \in \measm }}   \rho_{m, t},         \label{eq:objective-mp}    \\
    \text{subject to:}          
                                &~& \mathbf{h}(\mathbf{x}) = 0,                      \label{eq:h}            \\    
                                &~&  \mathbf{b}(\bm{\delta}) = 0,                     \label{eq:b}            \\
                                                                &~&  \bm{\rho}(\mathbf{y}) = 0,                     \label{eq:rho_bold}            \\
                                &~&  \mathbf{l}(\mathbf{x},\mathbf{y}, \bm{\delta}) \leq 0, \label{eq:l}            \\
                                &~&  \mathbf{k}(\mathbf{x}) \leq 0. \label{eq:k} 
\end{IEEEeqnarray}

Eq.~\eqref{eq:h} are the power flow equations, which consist of power balance~\eqref{eq:LD3Fkirchhoff} and generalized Ohm's law~\eqref{eq:LD3Fohm}.
\begin{equation}\label{eq:LD3Fkirchhoff}
    \sum_{k \in \loadsetmi \cup \gensetmi} \mathbf{s}_{k, t} = \sum_{ij \in \branchsetmi} \mathbf{s}_{ij, t} \; \; \forall i \in \busesm, t \in \timeseriesm,
\end{equation}
where $\mathbf{s}$ represents apparent power injections from the loads ($\loadsetmi$) and generators ($\gensetmi$) connected to bus $i$, and $\mathbf{s}_{ij}$ the apparent power flows to and from the edges connected to bus $i$ ($\branchsetmi$). The devices and branches connected to a bus are time invariant. In general, power terms relate as follows: $\mathbf{s} = \mathbf{p}+j\mathbf{q}$, with $j = \sqrt{-1}$, and $\mathbf{p},\mathbf{q} \in \mathbb{R}^{3 \times 1}$. 

The generalized Ohm's law for the LD3F formulation is derived in~\cite{Sankur} and reported here for the reader's comfort:
\begin{equation}\label{eq:LD3Fohm}
    \bm{\omega}_{i, t} = \bm{\omega}_{j, t} + \mathbf{A}_{ij} \cdot \mathbf{p}_{ij, t} + \mathbf{B}_{ij} \cdot \mathbf{q}_{ij, t} \; \; \forall (ij) \in \branchesm ,
\end{equation}
where
\begin{equation}
\mathbf{A}_{ij} = 2 (\mathfrak{Re}(\bm{\gamma}) \cdot \mathbf{R}_{ij} + \mathfrak{Im}(\bm{\gamma}) \cdot \mathbf{X}_{ij}) \; \; \forall (ij) \in \branchesm,
\end{equation}
\begin{equation}
 \mathbf{B}_{ij} = 2 (\mathfrak{Re}(\bm{\gamma}) \cdot \mathbf{X}_{ij} - \mathfrak{Im}(\bm{\gamma}) \cdot \mathbf{R}_{ij}) \; \; \forall (ij) \in \branchesm.
\end{equation}
All impedance-related terms are time invariant, the resistance $\mathbf{R}_{ij}$ and reactance $\mathbf{X}_{ij}$ are real $3 \times 3$ matrices, and $\bm{\gamma}$ is:
\begin{equation}
\bm{\gamma} = \begin{bmatrix} 
1 & \alpha^2 & \alpha \\
\alpha & 1  & \alpha^2 \\
\alpha^2 & \alpha & 1
\end{bmatrix},
\end{equation}
with $\alpha = e^{-j 2/3 \pi}$.

Eq.~\eqref{eq:b}-\eqref{eq:l} are the fulcrum of the MILP-SE-PI, as they contain the binary variables and relate them to the rest of the problem. These equations can differ for single- and three-phase users, and are introduced here for the former case first.

Note that in this paper we assume that measurements are only available at end-user locations. As such, all other buses are modelled as zero-injection buses (as in standard SE theory), which are naturally incorporated in the MILP-SE-PI thanks to~\eqref{eq:h}. Nevertheless, if additional measurements are available, they can be included, as discussed later in this section. Finally,~\eqref{eq:k} indicates any upper/lower bounds that can be applied to the system variables. The better the a-priori knowledge of the network and user behaviour, the tighter the bounds can be.

\subsection{Single-phase Users}

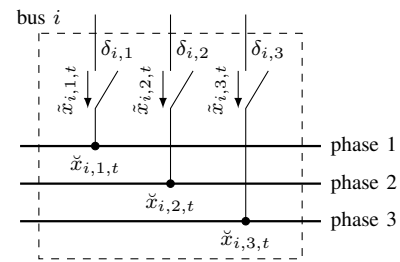
\begin{figure}[b]
    \centering
    \begin{tikzpicture}
        \draw [line width = 0.3mm] (0.0, 0.0) -- (4.0, 0.0) node [right] {\footnotesize phase 1};
        \draw [line width = 0.3mm] (0.0,-0.5) -- (4.0,-0.5) node [right] {\footnotesize phase 2};
        \draw [line width = 0.3mm] (0.0,-1.0) -- (4.0,-1.0) node [right] {\footnotesize phase 3};
        \draw (1.0,1.75) -- (1.0,1.0); 
        \draw [-latex] (0.9,1.0) -- (0.9,0.5) node [pos=0.5, above, rotate=90] {\footnotesize $\edgevarm_{i,1,t}$};
        \draw (1.3,1.0) node [above] {\footnotesize $\delta_{i,1}$} -- (1.0,0.5) -- (1.0, 0.0) node [below] {\footnotesize $\nodevarm_{i,1,t}$}; 
        \draw [fill] (1.0, 0.0) circle (0.5mm);
        \draw (2.0,1.75) -- (2.0,1.0); 
        \draw [-latex] (1.9,1.0) -- (1.9,0.5) node [pos=0.5, above, rotate=90] {\footnotesize $\edgevarm_{i,2,t}$};
        \draw (2.3,1.0) node [above] {\footnotesize $\delta_{i,2}$} -- (2.0,0.5) -- (2.0,-0.5) node [below] {\footnotesize $\nodevarm_{i,2,t}$}; 
        \draw [fill] (2.0,-0.5) circle (0.5mm);
        \draw (3.0,1.75) -- (3.0,1.0); 
        \draw [-latex] (2.9,1.0) -- (2.9,0.5) node [pos=0.5, above, rotate=90] {\footnotesize $\edgevarm_{i,3,t}$};
        \draw (3.3,1.0) node [above] {\footnotesize $\delta_{i,3}$} -- (3.0,0.5) -- (3.0,-1.0) node [below] {\footnotesize $\nodevarm_{i,3,t}$}; 
        \draw [fill] (3.0,-1.0) circle (0.5mm);
        \draw [dashed] (0.25,1.5) node [above] {\footnotesize bus~$i$} rectangle (3.75,-1.5);
    \end{tikzpicture}
    \caption{Model of a single phase user, at a given time step~$t$.}
    \label{fig:phase-id}
\end{figure}

Fig.~\ref{fig:phase-id} illustrates the concept: single-phase connections are modelled as three-phase, but to each phase $\phi \in \Phi$ a binary variable $\delta_{i,\phi}$ is assigned that determines whether that phase is actually the connected one ($\delta_{i,\phi} = 1$) or not ($\delta_{i,\phi} =0$). The following constraint enforces single-phase connectivity: 
\begin{IEEEeqnarray}{ l l }\label{eq:deltasum1p}
    \sum_{\phi \in \Phi} \delta_{i,\phi} = 1,
        &\quad \forall i \in \singlephaseusersetm,
\end{IEEEeqnarray}
where $\singlephaseusersetm$ is the set of single-phase users. Note that \eqref{eq:deltasum1p} is part of \eqref{eq:b} and is time-invariant: it is assumed that the phase connectivity does not change during the measurement collection period.

Let the (overall) measurement set $\measm$ be defined as $\measm = \mathcal{M}^{u} \cup \mathcal{M}^{o}$. $\mathcal{M}^{u}$ is the set of measurements from the devices of users with unknown connectivity. $\mathcal{M}^o$ indicates any other available measurements, e.g., transformer power supply, power flows in other cables, etc.
Let us further divide $\measm^u$ depending on the measured quantity, e.g., power injection. These quantities are part of the $\mathbf{x}$ variables and can be divided in two categories: 
\begin{enumerate}
    \item node variables:~\nodevar, 
    \item edge variables:~\edgevar. 
\end{enumerate}
In this paper, \nodevar \ refers to $\omega$ and \edgevar \ to $p$ and $q$, but should another formulation/measurement set-up be used, \nodevar \ generalizes to any voltage measurements, e.g., magnitude, phasor, and \edgevar \ to any current or power measurements. The difference between \nodevar \ and \edgevar \ is that in non-connected phases, the latter can be set to zero: there is no flow/injection. Conversely, the voltage variables of non-connected phases, cannot be set to zero. These will actually take the same value as the respective phase in the adjacent bus (i.e., the bus at the three-phase end of the service cable), which is also a problem variable.

Let the measurements that refer to node and edge variables be called ``node measurements" and ``edge measurements", and let their sets be $\nodemeasm, \edgemeasm$, such that $\measm^u = \nodemeasm \cup \edgemeasm $.

\subsubsection{Single-phase Users - Node Measurements}

Each measurement $m \in \nodemeasm$ relates to a variable $x_{m, t} \in \mathbf{x}$, and is described by the measurement value $z_{m, t} \in \mathbb{R}$, as reported by the SM \footnote{Single-phase users are equipped with single-phase SM, hence $z_{m,t}$ is a scalar.}, and some confidence on the quality of the measurement, given by the SM specifications: $\sigma_{m} \in \mathbb{R}$ ($\sigma$'s are assumed time-invariant). Every user $u \in \singlephaseusersetm$ is connected to a bus~$i \in \busesm^{1p} \subset \busesm$, and a mapping exists between measurement~$m$ and bus~$i$ (indicated as $m \rightarrow i$, or $i \rightarrow m$ where more convenient), so that these two indices can be used interchangeably. For any node measurement, the following hold:
\begin{IEEEeqnarray}{ l l }
    &\quad -a_{i,\phi} \cdot (1-\delta_{i,\phi}) \leq y_{m,t} - \nodevarm_{i,\phi,t} \leq  b_{i,\phi} \cdot (1-\delta_{i,\phi}),     \nonumber \\
        &\quad \hspace{1.5cm} \forall m \in \nodemeasm, \phi \in \Phi, t \in \timeseriesm: m \to i \in \busesm^{1p},   
        \label{eq:node_constraint} \\
    &\quad \rho_{m,t} = f(y_{m,t}),                                          
     \forall m \in \nodemeasm, t \in \timeseriesm  \label{eq:first_paper_residual}. 
\end{IEEEeqnarray}

Eq.~\eqref{eq:node_constraint} is part of \eqref{eq:l} and ensures that if a user is connected to a specific phase~$\phi$ of a bus~$i$, the corresponding node variable~$\nodevarm_{i,\phi,t}$ equals the auxiliary variable~$y_{m,t}$. Eq.~\eqref{eq:first_paper_residual} indicates that the residual for $m$ is assigned via the auxiliary variable~$y_{m,t} \in \mathbf{y}$, contrary to standard SE, where the residual would be a function of $\nodevarm_{i, \phi, t}$. Note that the residuals are part of the objective~\eqref{eq:objective-mp}, and the $f$ used in \eqref{eq:first_paper_residual} in this paper is described in~\eqref{eq:WLAV}. Thus, the introduction of $y_{m,t}$ allows to only have one residual equation per measurement, instead of three. This makes the MILP more efficient because it reduces the number of terms in the objective, leaving out terms that correspond to variables that have no physical meaning\footnote{i.e., the $x$'s of non-connected phases.} and whose value could oscillate, affecting convergence and accuracy.  $a_{i,\phi}, b_{i,\phi} \in \mathbb{R}$ are generic bounds and can be bus-, time- and/or phase-variant or not. A practical assignment is $a_{i,\phi} = b_{i,\phi} = \overline{\nodevarm_{i,\phi}}$, where $\overline{\nodevarm_{i,\phi}}$ is the variable's upper bound.  

A residual describes some type of ``distance" between a variable's most-likely value and its measured value. In SE, the most popular residual definition is the quadratic weighted least squares~\cite{Vanin2021}. In this paper, to avoid non-linearities, weighted least absolute values (WLAV) are used instead:
\begin{equation}\label{eq:WLAV}
\rho_{m, t} =  | y_{m, t} - z_{m, t} |/\sigma_m,\quad \forall m \in \measm^u, t \in \timeseriesm,
\end{equation}
which is implemented using its exact linear relaxation:
\begin{IEEEeqnarray}{rCl}
     \rho_{m, t} \geq \frac{y_{m,t} - z_{m,t}}{\sigma_m},\quad \forall m \in \measm^u, t \in \timeseriesm \label{eq:rWLAV1}\\
     \rho_{m, t} \geq - \frac{y_{m,t} - z_{m, t}}{\sigma_m},\quad  \forall m \in \measm^u, t \in \timeseriesm \label{eq:rWLAV2}.
\end{IEEEeqnarray}
Eq.~\eqref{eq:WLAV} is part of \eqref{eq:rho_bold} and is the same for node and edge measurements.

\subsubsection{Single Phase Users - Edge Measurements}

Edge measurements are described by the \edgevar \ equivalent of \eqref{eq:node_constraint}-\eqref{eq:WLAV} and the following additional constraint:
\begin{equation}
     \underline{\edgevarm_{i,\phi}} \cdot \delta_{i,\phi} \leq \edgevarm_{i,\phi,t} \leq \overline{\edgevarm_{i,\phi}} \cdot \delta_{i,\phi}, \; \; \forall i \rightarrow m \in \edgemeasm, \phi \in \Phi, t \in \timeseriesm.                     
        \label{eq:edge_constraint}   
\end{equation}
Eq.~\eqref{eq:edge_constraint} ensures that if the user is not connected to~$\phi$, the corresponding edge variable~$\edgevarm_{i,\phi,t}$ equals zero. If it is connected, lower ($\underline{\edgevarm_{i,\phi}}$) and upper ($\overline{\edgevarm_{i,\phi}}$) bounds are given. These can stem from knowledge on the user's connected capacity or other system information. As with all other bounds, the better the system knowledge, the tighter they can be, but the modeller should pay attention to not assign too tight bounds that exclude optimal solutions. 

\subsection{Three-phase Users}

Three techniques to model three-phase users are considered:
\begin{itemize}
    \item Model A: Like three separate single-phase users, using the equations presented so far,
    \item Model B: Like model A, but adding a constraint \eqref{eq:model_b} to enforce that each of the three resulting single-phase users is connected to a different phase, 
    \item Model C: a ``connection mapping" that reduces the number of binary variables.
\end{itemize}

In Model B, for every three-phase user $u$ that is ``split" in three single-phase users $k, l, m$, the following is added to \eqref{eq:b}:
\begin{equation}
    \sum_{j \in \{k, l, m\}} \delta_{j, \phi} = 1 \; \; \; \forall \phi \in \Phi.
    \label{eq:model_b}
\end{equation}

While the implementation of Model A and B is quite straightforward, the total number of binary variables associated to the three-phase user is $3 \times 3 = 9$: three for every separate single-phase user. With Model C, these are reduced to 6, which reduces the size of the branch-and-bound/cut search tree. Note that three-phase users have three-phase SM, and, as such, one measurement of every (edge or node) quantity per phase. In models A and B, each vector is ``split" to be compatible with a single-phase model. For model C this is not necessary.

\subsubsection{Model C - Node Measurements}
\begin{figure}[b]
\centering
\includegraphics[width=.4\textwidth]{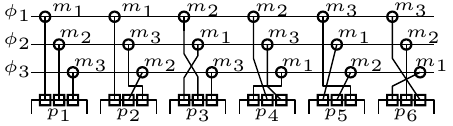}
\caption{Illustration of the three-phase user mapping, where the bottom squares indicate the SM channels.}
\label{fig:mapping}
\end{figure}

Let each three-phase user $u \in \threephaseusersetm $ connected to bus~$ i \in \busesm^{3p} \subset \busesm $ have a measurement vector~$\bar{m}~\in~\mathbb{R}^{3~\times~1}$, and let a mapping exists between~$\bar{m}$ and bus~$i$ like that for single-phase users. There are six possible ways~$p \in \Delta$ of connecting a three-phase user to a three-phase bus, each of which maps to a specific permutation of the phase-sequence:~$p \to \bar{\phi}$~(Table~\ref{tab:connection}, Fig.~\ref{fig:mapping}). The selection of a connection sequence~$p \in \Delta$ is described by a binary variable~$\delta_{i,p}$. If sequence $p$ is chosen, $\delta_{i,p} = 1$, else $\delta_{i,p} = 0$. Any three-phase node measurement is described by:
\begin{IEEEeqnarray}{l l}
&\quad \rho_{\bar{m},t} = f(y_{\bar{m},t}),  \forall \bar{m} \in \nodemeasm, t \in \timeseriesm, \nonumber
        \\
&\quad -a_{i,\phi} \cdot (1-\delta_{i,p}) \leq y_{\bar{m},t} - \nodevarm_{i,\bar{\phi},t} \leq  b_{i,\phi} \cdot (1-\delta_{i,p}), \hspace{1cm} \label{eq:node_constraint_3f}\\ 
&\quad \hspace{1cm} \forall \bar{m} \in \nodemeasm, p \in \Delta, t \in \timeseriesm: \bar{m} \to i \in \busesm^{3p}, \delta \to \bar{\phi}. \nonumber   
\end{IEEEeqnarray}
Constraint~\eqref{eq:node_constraint_3f} is the three-phase equivalent of~\eqref{eq:node_constraint}. 

\subsubsection{Model C - Edge Measurements}

A three-phase edge measurement is described by:
\begin{IEEEeqnarray}{ l l }
&\quad \rho_{\bar{m},t} = f(y_{\bar{m},t}), 
        \forall \bar{m} \in \edgemeasm, t \in \timeseriesm, \hspace{2cm}
        \nonumber \\
&\quad  - a \cdot (1-\delta_{i,p}) \leq y_{\bar{m},t} - \edgevarm_{i,\bar{\phi},t} \leq  b \cdot (1-\delta_{i,p}),  \nonumber \\   
        &\quad \hspace{0.5cm} \forall \bar{m} \in \edgemeasm, p \in \Delta, t \in \timeseriesm: \bar{m} \to i \in \busesm^{3p}, \delta \to \bar{\phi}. 
        \label{eq:edge_constraint_3f_1}
        \\
&\quad    \underline{\edgevarm_{i,\phi}} \cdot \sum_{p \in \Delta} \delta_{i,p} \leq \edgevarm_{i,\phi,t} \leq \overline{\edgevarm_{i,\phi}} \cdot \sum_{p \in \Delta} \delta_{i,p}, \nonumber \\
        &\quad \hspace{3cm} \forall i \in \busesm^{3p}, \phi \in \Phi, t \in \timeseriesm.
        \label{eq:edge_constraint_3f_2}
\end{IEEEeqnarray}

\begin{table}[b]
    \centering
    \caption{Mapping for three-phase households:~$p \to \bar{\phi}$}
    \label{tab:connection}
    \begin{tabular}{c | c c c c c c }
                & $p_{1}$  & $p_{2}$  & $p_{3}$  & $p_{4}$  & $p_{5}$  & $p_{6}$  \\ \hline
        $m_{1}$ & $\phi_{1}$    & $\phi_{1}$    & $\phi_{2}$    & $\phi_{2}$    & $\phi_{3}$    & $\phi_{3}$    \\
        $m_{2}$ & $\phi_{2}$    & $\phi_{3}$    & $\phi_{1}$    & $\phi_{3}$    & $\phi_{1}$    & $\phi_{2}$    \\
        $m_{3}$ & $\phi_{3}$    & $\phi_{2}$    & $\phi_{3}$    & $\phi_{1}$    & $\phi_{2}$    & $\phi_{1}$
    \end{tabular}
\end{table}

Lastly, the following constraint ensures that the user is connected through exactly one connection sequence:
\begin{IEEEeqnarray}{ l l }\label{eq:deltasum3p}
    \sum_{p \in \Delta} \delta_{i, p} = 1,
        &\quad \forall i \in \threephaseusersetm.    
\end{IEEEeqnarray}

\subsection{Other Measurements}

If other measurements $\measm^o$ are available, which are not associated to a user whose connectivity needs to be determined, such as transformer measurements, these can be included in \eqref{eq:objective-mp} in a ``normal" WLAV-SE fashion:
\begin{IEEEeqnarray}{rCl}
     \rho_{m, t} \geq \frac{x_{m,t} - z_{m,t}}{\sigma_m},\quad \forall m \in \measm^o, t \in \timeseriesm \label{eq:rWLAV1-transfo}\\
     \rho_{m, t} \geq - \frac{x_{m,t} - z_{m, t}}{\sigma_m},\quad  \forall m \in \measm^o, t \in \timeseriesm \label{eq:rWLAV2-transfo}.
\end{IEEEeqnarray}
No auxiliary variables are needed and there is no difference between edge and node measurements. If the connectivity of a subset of the users is known for some reason, \eqref{eq:rWLAV1-transfo}-\eqref{eq:rWLAV2-transfo} can be applied to their measurements, too.

\subsection{Notes on Implementation and Measurement Choice}

The MILP-SE-PI is implemented in Julia/JuMP~\cite{JuMP}, and makes use of some functions from the PowerModelsDistribution.jl~\cite{PMD_PSCC} and PowerModelsDistributionStateEstimation.jl~\cite{Vanin2021} libraries. These are flexible toolboxes which make it relatively easy to choose different power flow formulations or measurement types, like phasors. In this paper, the measurements used are fifteen-minute averages of active and reactive power and voltage magnitude. These match the present capabilities of Flemish SMs, and countries that adopt a similar technology. Voltage magnitude measurements are squared to match the square voltage ($\bm{\omega}$) variables in~\eqref{eq:node_constraint}-\eqref{eq:node_constraint_3f}. Users can both be consuming or injecting power from/to the grid. Only user SM measurements are assumed available. Furthermore, we assume that the phases of \textit{all} users need to be determined\footnote{This is a worst-case scenario: if the connectivity of a user is known, this can simply be fixed, reducing the number of binaries.}. Given the last two statements, we get that $\measm^o = \emptyset$. 

In general, the proposed method is sound with any combination of measurements as long as the underlying continuous problem is overdetermined, i.e., the system is ``observable". Furthermore, the SE by design ``filters" measurement noise, reducing the impact of measurement errors on the PI process. 
%
\section{Improving the Computational Time}\label{sec:improvements}

This section describes several techniques that have been explored to achieve faster MILP solutions than with a ``vanilla" implementation of the model in Section~\ref{sec:mathematical_model}. Useful insights have been found in paper~\cite{KlotzMILP}. Observing the MILP-SE-PI response to these techniques gives interesting information on the problem's behaviour. 
First of all, the underlying LP is prone to poor scaling, as it is often the case for SE in DNs. Nevertheless, an attentive choice of per-unit values and other coefficients improved the solve times without changing the results. Secondly, we tested the MILP-SE-PI behaviour with different MIP starts, in which increasing subsets of the users are assigned to the correct phase. This did not significantly improve (nor worsen) the solve speed: the solver spends most time proving the optimality of the solution rather than finding a good/the optimal one\footnote{This means that even if the solve time exceeds a pre-established (reasonable) cut-off time, the hitherto obtained best solution is likely accurate.}. Some additional considerations on MIP starts are made in Section~\ref{sec:result_limited_eulv}. Furthermore, Benders decomposition (BD) was tried, as this was also the technique of choice to solve the MILP-PI in~\cite{Heidari-Akhijahani2021}, although the decomposition is different due to the different problem formulation. All constraints are time dependent except~\eqref{eq:b}. $\bm{\delta}$ are so-called complicating variables, and BD can be applied to a MILP with this structure. 
However, as also observed in~\cite{Heidari-Akhijahani2021}, BD alone does not solve in acceptable time. In~\cite{Heidari-Akhijahani2021}, this is tackled with a BD acceleration technique. In this paper, a more reliable network decomposition is applied to larger feeders instead. This is discussed hereafter, together with a technique to tighten the problem formulation. 




\subsection{Tightening the Formulation}\label{sec:tightening}

The MILP formulation~\eqref{eq:objective-mp}-\eqref{eq:k} can be tightened if information on the overall user distribution per phase is available. Then, the following can be added to~\eqref{eq:b}:

\begin{equation}
 \underline{N} \leq \sum_{i \in \busesm} \delta_{i, \phi} \leq \overline{N} \; \; \forall \phi \in \Phi,
\end{equation}

where $\overline{N}, \underline{N} \in \mathbb{R}$ are a maximum and minimum number of expected user connections per phase. For instance, for a WYE-connected feeder that has 10 users and non-zero power measurements in all the transformer's phases, $\overline{N}~=~8, \underline{N}~=~1$ can be safely assumed. Better/tighter values can be guessed by examining the measurement data prior to starting the MILP, and have a positive impact on the computational time.

\subsection{Network Decomposition}\label{sec:network-decomposition}

\begin{figure*}
    \centering
    \begin{tabular}{cc}
  \includegraphics[width=0.49\textwidth]{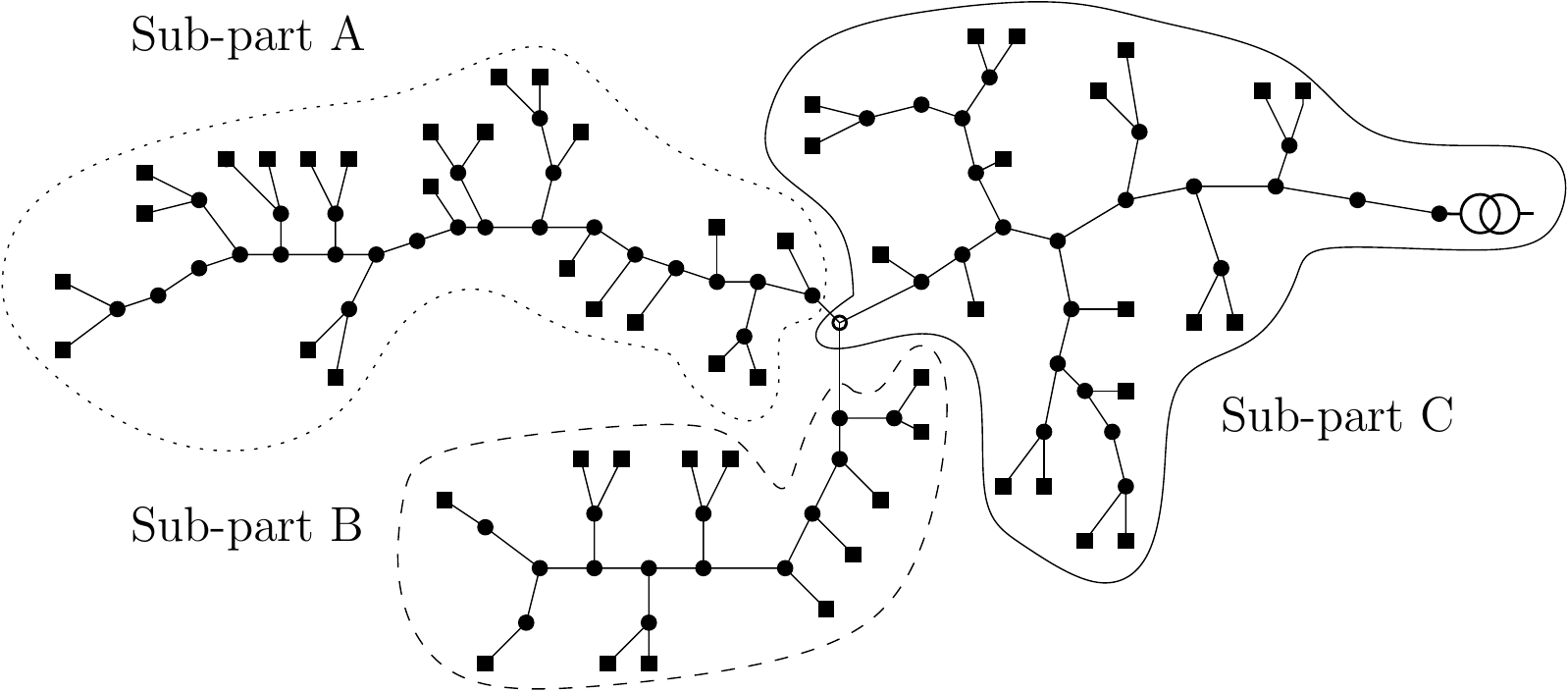} &
  \includegraphics[width=0.49\textwidth]{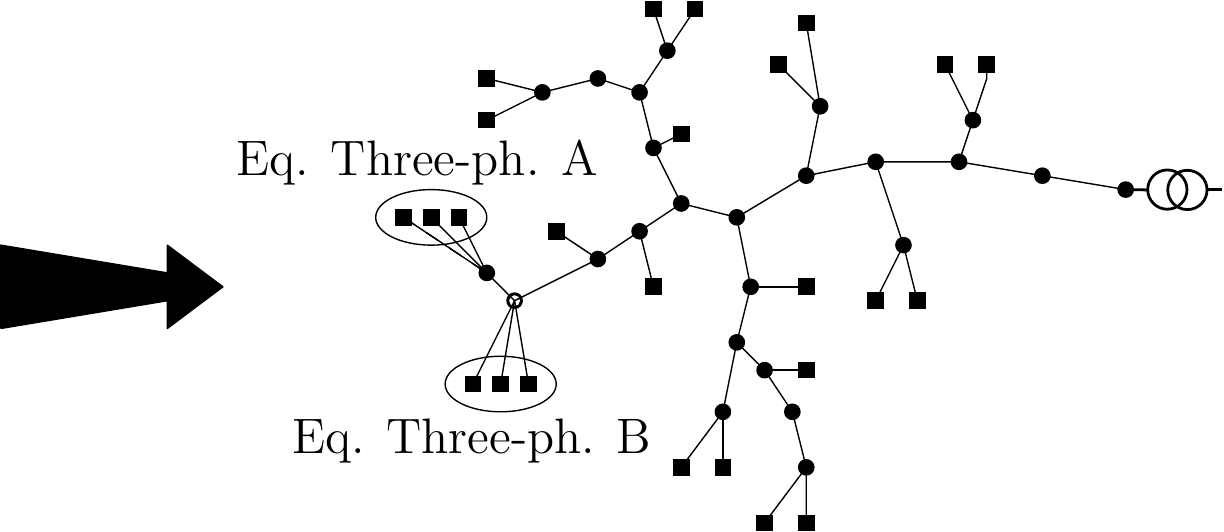} \\
\end{tabular}
    \caption{Network decomposition applied to the IEEE European LV Feeder. Phase connections are first identified for the ``extreme" sub-parts A and B. These are then re-united to Sub-part C as equivalent three-phase users. Squares indicate buses to which users are connected, with the exception of the equivalent users in the right part of the picture, which are connected to the buses where the network was split.}%
    \label{fig:EULVparts}%
\end{figure*}

The vast majority of the distribution networks are operated radially. As such, feeders are tree (or chain) graphs and they can be divided in sub-parts, as illustrated in Fig.~\ref{fig:EULVparts} for the IEEE European Low Voltage Test Feeder. PI can be performed for Sub-part A and Sub-part B first, and then these are re-united to Sub-part C in the form of equivalent three-phase users. When solving Sub-part A and Sub-part B, the node by which they were originally
connected to the rest of the tree is modelled as a slack-bus. If measurements are available for that node, they can be included. However, this seems unlikely in LVDNs (as it is a three-phase bus on the main feeder cable), so here no measurements are assumed. The equivalent users are provided with ``artificial" three-phase voltage magnitude and active and reactive power measurements that correspond to the power supply and voltage of the Sub-parts' slack buses. These are inherently calculated by the SE jointly to the sub-part phases.

This decomposition is effective because, from a SE and PI standpoint, Sub-parts A and B are as observable as the full network: they preserve the same measurement set-up (end-user-only measurements), and ``upstream" power flows do not affect the SE/PI of trees that are further away from the root bus. In this exercise, the sub-trees have been chosen by visual inspection. To scale up the application of this method, algorithms could be put in place, that automatically split the network ensuring that all sub-trees have a sufficiently small number of nodes.




\section{Case Studies}\label{sec:case_studies}

\subsection{Description}

The case studies are inspired from those in~\cite{Heidari-Akhijahani2021} as, to the best of the authors' knowledge,~\cite{Heidari-Akhijahani2021} is the only other reference that addresses PI by modelling the feeders with power flow equations. As such, our method shares with~\cite{Heidari-Akhijahani2021} the susceptibility to errors in the underlying network model, addressed in this section. However, a rigorous comparison between~\cite{Heidari-Akhijahani2021} and the presented results is not possible, as the user demand/generation profiles are different (whereas the (largest) test feeders used are the same). For benchmarking purposes, researchers could benefit from a structured library with both network and profile data, similar to that for optimal power flow~\cite{pglib}, but this currently does not exist for PI.

Network and profile data are taken from the ENWL database~\cite{ENWL}, from which network 1, feeder 1 (i.e. the IEEE European LV Test Feeder) and network 5 feeder 4 are chosen. These have, respectively, 55 and 30 single-phase users, and 906 and 457 buses. The number of buses has been reduced to 118 and 65, respectively, with the method described in~\cite{Claeys2021}. The networks after this pre-processing step are electrically equivalent to the original ones, no approximations or information losses are introduced. Furthermore, network 4 feeder 5 and network 11 feeder 2 are used to compare different three-phase user models.

The profile data from~\cite{ENWL} have five-minute resolution. To create ``synthetic" measurements, \textit{ 1)} power flow calculations are run with the five-minute power data, then \textit{ 2)} Gaussian errors are added to both the original profiles and the voltage magnitudes resulting from the power flow. Finally, \textit{3)} the average measurement values of three adjacent timesteps are taken, to simulate fifteen-minute measurements. The five-minute power profiles in the database (and subsequently the resulting voltage magnitudes) vary over time, but the variations are less sharp than by instantaneous and high-resolution measurements. While more prone to errors due to loss of synchronism, instantaneous and high resolution measurements would contain more information for PI (and other parameter estimation) purposes. However, fifteen-minute averages is currently the default resolution of several commercial SMs~\cite{Raggi2022}, including those used in Flanders, and hence has been used as set-up in this paper.  

The measured values $z_m$ appear in~\eqref{eq:objective-mp}, while $\sigma_m$ are the standard deviation of the Gaussian distribution used to create the errors, and correspond to a third of the maximum measurement error given by the SM class. The following maximum SM errors are considered: 0.5\%, 1\%, 5\%. Furthermore, to simulate inaccurate network data, errors are also added to cable impedance values, with $3 \sigma_m$ set to 0\%, 10\%, 20\% and 50\%, like in~\cite{Heidari-Akhijahani2021}. It is assumed that every user has a SM with fifteen-minute average voltage magnitude and average active and reactive power measurements, and that the same measurements are available at the LV side of the transformer as well. In the authors' experience, such a set-up is reasonable in the field. As mentioned before, the proposed method is based on SE and would work with any measurement combination, as long as the system is observable. This means that transformer measurements, if unavailable, are not necessary if there are enough SMs. However, without a reference measurement for the feeder head, users would be assigned to different phases, but it would be impossible to assess how these phases relate to the transformer's, leaving the network data incomplete. This applies to any PI method, not only the proposed MILP-SE-PI.

 \begin{table*}[b]
           \centering
           \caption{PI accuracy for the 30-user feeder, values in cells are percentages.}
           \begin{tabular}{l|l||c|c|c|c|c||c|c|c|c|c||c|c|c|c|c}
           \hline

           \multicolumn{2}{c ||}{SM Error $\rightarrow$}&\multicolumn{5}{c||}{0.5\%}&\multicolumn{5}{c||}{1\%}&\multicolumn{5}{c}{5\%} \\

           \hline

           \multicolumn{2}{c|}{}&\multicolumn{15}{c}{Meas. Time Length (\#time steps)} \\
          
           \hline
          
          \multicolumn{2}{c|}{Impedance err. $\downarrow$} & 5 & 10 & 20 & 40 & 60 & 5 & 10 & 20 & 40 & 60 & 5 & 10 & 20 & 40 & 60 \\
          \hline
          \hline
          
\multirow{2}{*}{MILP} & 0\% & 100.0 & 100.0 & 100.0 & 100.0 & 100.0 & 100.0 & 100.0 & 100.0 & 100.0 & 100.0 & 88.0 & 94.0 & 100.0 & 100.0 & 100.0 \\
\multirow{2}{*}{SE-PI} & 10\% & 100.0 & 100.0 & 100.0 & 100.0 & 100.0 & 100.0 & 100.0 & 100.0 & 100.0 & 100.0 & 88.0 & 94.0 & 100.0 & 100.0 & 100.0 \\
& 20\% & 100.0 & 100.0 & 100.0 & 100.0 & 100.0 & 100.0 & 100.0 & 100.0 & 100.0 & 100.0 & 88.0 & 94.0 & 100.0 & 100.0 & 100.0  \\
& 50\% & 100.0 & 100.0 & 100.0 & 100.0 & 100.0 & 100.0 & 100.0 & 100.0 & 100.0 & 100.0 & 88.0 & 94.0 & 100.0 & 100.0 & 100.0  \\
\hline
Corr. & - & 51.3 & 44.0 & 20.3 & 50.6 & 67.3 & 43.3 & 33.0 & 16.6 & 41.3 & 51.0 & 36.3 & 27.0 & 23.3 & 33.3 & 35.3 \\
\hline
Regr. & - & 33.6 & 41.0 & 31.0 & 51.3 & 58.7 & 32.0 & 32.7 & 28.7 & 38.0 & 34.7 & 34.7 & 34.7 & 29.0 & 31.3 & 31.3 \\
\hline
\end{tabular}\label{tab:results_30loads}
\end{table*}

For underdetermined systems, other methods than the proposed one might be preferable. In particular, voltage-based methods exist that can determine the connectivity of a few SMs when most of the system is not monitored~\cite{Hoogsteyn2022}. However, PI cannot be performed for non-monitored connections, so a full picture of the system cannot be obtained, which has a negative impact on any calculation reliant on network data. 

The results of the proposed method are compared to those of two techniques from the literature: Pearson correlation and linear regression. These methods are - by construction - not affected by errors in the network impedances, and are also used as a benchmark in~\cite{Heidari-Akhijahani2021}. The first relies on voltage measurements only: Pearson correlation is calculated between the voltage measurement time series of each user and each transformer phase~\cite{PezeshkiAUPEC, Hoogsteyn2022}. The user is then assigned to the transformer phase with the highest correlation. The linear regression method uses both voltage magnitude and active power measurements, as in~\cite{Short2013}. 

The next part of this section shows the PI accuracy of the three methods and the computational time of the proposed MILP-SE-PI. The accuracy is defined as the percentage of users for which the phase connectivity is correctly identified. Computational times of Pearson correlation and linear regression are not shown as both are negligible compared to the MILP-SE-PI.

The calculations are performed on a 64-bit linux server with Intel Xeon Platinum 8268 CPU @ 2.90GHz. Gurobi version 9.1 is accessed via Gurobi.jl and solves the MILP-SE-PI, with the following parameters: ``Threads" = 30, ``Presparsify" = 1, ``Cuts" = 3,``MIP\_focus" = 3~\cite{Gurobi}. All other parameters keep their default values, which means that the MIP optimality gap is 1E-4. For clarity, we do not perform any relaxation of the MILP prior to solving it with Gurobi.

The power flow calculations to generate the measurement data are run on an exact, non-convex power flow formulation and solved with Ipopt~\cite{ipopt}. All the problem bounds are set to be much larger than nominal values (e.g., $\overline{\edgevarm_{i, \phi}}$ in \eqref{eq:edge_constraint} is 10 times the transformer nominal value), and the tightening technique in~\ref{sec:tightening} is not used. This avoids strong assumptions on the network behaviour, and might lead to longer solve times but circumvents the risk of erroneous results, and simulates a case in which the modeller has very limited system information.

\subsection{Numerical Results: 30 Single-phase User Feeder}\label{sec:results_30-user_feeder}

For this feeder, the decomposition method described in~\ref{sec:network-decomposition} is not necessary. While we do not have full data for other countries, it is interesting to mention that about 84\% of Flemish feeders have fewer than 30 users, and as such can reasonably be expected to be solved for with the MILP-SE-PI method without the decomposition step. All the percentage values in the presented tables are average values over 10 calculations with 10 different random error assignments, unless stated otherwise. For this feeder, results are reported in Table~\ref{tab:results_30loads}. It can be shown that for the lower SM errors, 100\% accuracy is achieved with very limited time series data. This suggests that the LD3F does not introduce significant modelling errors, which was expected given the encouraging results in~\cite{Vanin2020}. Furthermore, the results indicate that impedance errors do not jeopardize the accuracy of the method, whereas more than 10 measurement time steps are required to compensate for the noise introduced by larger SM errors (5\%). The accuracy of the MILP-SE-PI outperforms that of the benchmarking methods.

The reason why impedance errors do not compromise the PI accuracy is that topological errors (like phase connectivity) have in general a much larger impact on the measurement residuals. This has been discussed in Chapter 8.3 of~\cite{bible} and in~\cite{VaninImpedance}, where high loading conditions are required to observe a significant residual impact of impedance errors in the European LV Feeder. 

\begin{figure}[b]
\includegraphics[width=.45\textwidth]{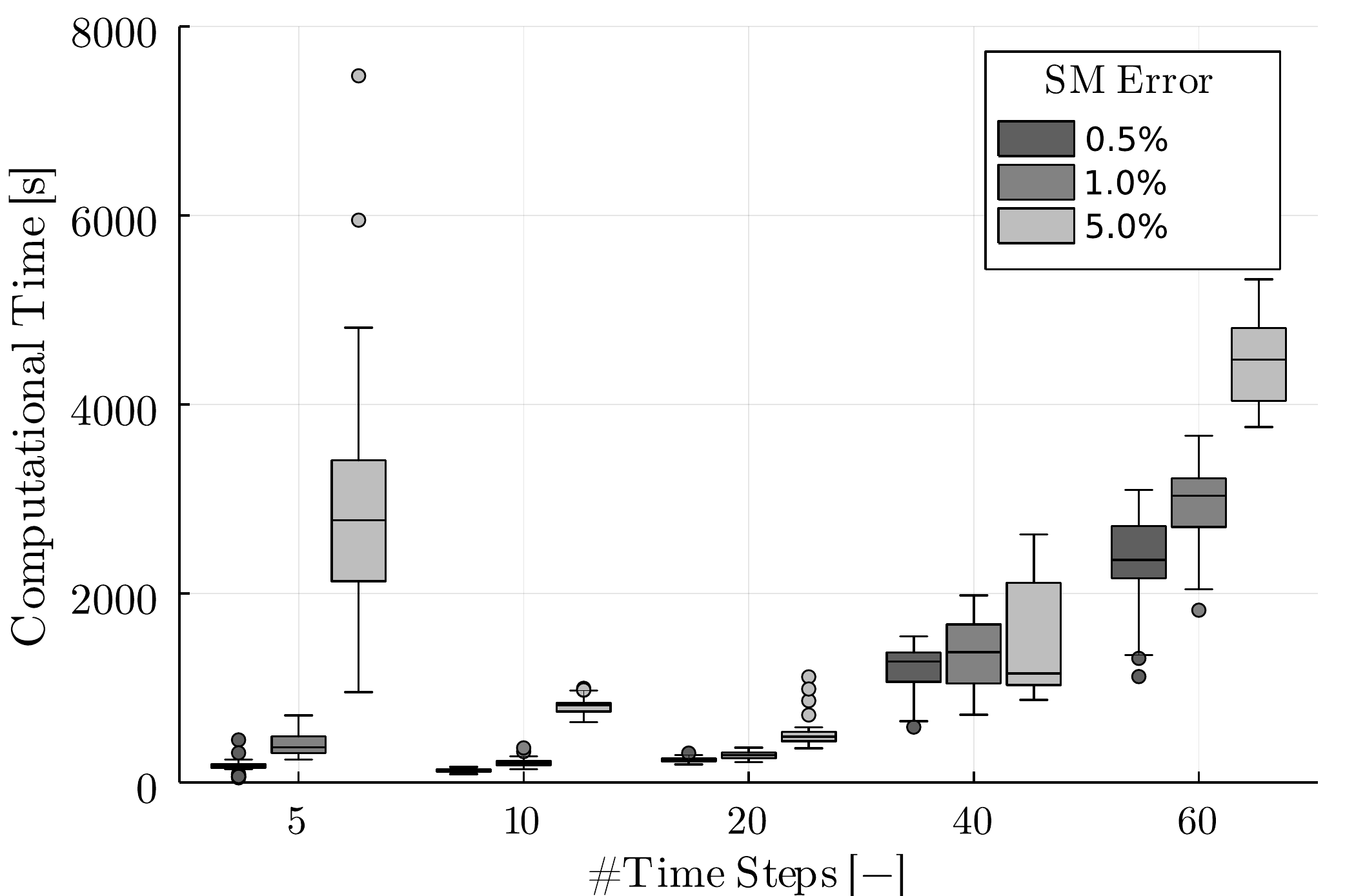}
\caption{MILP-SE-PI solver time for the 30-users feeder.}
\label{fig:solve_time_30l}
\end{figure}

The computational time required to solve the MILP-SE-PI can be seen in Fig.~\ref{fig:solve_time_30l}. Except for the case of 5\% SM errors with 5 time steps, the general trend is that longer measurement time series require longer solve times. This is expected: while the binary variables are time-independent, the number of continuous system variables increases with the number of time steps, increasing the computational effort. Increasing SM errors also appear challenging for the solver, and lead to extra computational efforts, whereas increasing impedance errors does not. 54\% of the overall instances solve in less than 15 minutes, and 91\% in less than one hour. Note that the extra measurement acquisition time required by the other methods to obtain comparable accuracies highly exceeds the MILP-SE-PI solve time. For example, in the 0.5\% SM error case, Pearson correlation requires 50 time steps to achieve 90\% accuracy, while 5 time steps are sufficient with the MILP-SE-PI. The difference of 45 quarter-hour measurement time steps translates to 11.25 h.

\subsection{European LV Feeder - Without Decomposition}\label{sec:result_limited_eulv}

In this section, results for the European LV Feeder are presented, which are summarized in Table~\ref{tab:partial_results_55loads}. These results refer to calculations with no error on impedance data and only for the case of 1\% SM error. The reason for addressing only this case is that, without the decomposition in Section~\ref{sec:network-decomposition}, calculating a single PI instance for this feeder exceeds the cut-off limit of 2.5 hours imposed to the solver. As such, repeating the numerical exercise for all the cases of Section~\ref{sec:results_30-user_feeder} would take excessively long. Nevertheless, it is worth reporting that, even though the solver does not explore the whole search tree in the allowed solve time, the MILP-SE-PI still outperforms the other methods' accuracy (see Table~\ref{tab:partial_results_55loads}). Cutting off before the end of the solving process implies that the optimality of the current best MILP solution is not proven, but this is also the case for any other heuristic or statistical method. As a matter of fact, since branch-and-bound/cut gradually improves on the previous found solution, a possible approach could be to first calculate a quick solution with a fast method, and provide that as a MIP start to improve on for a given calculation time. Finally, in real-life, the calculations would only need to be performed once, so the cut-off time could be increased significantly, possibly solving in acceptable time also for feeders of this size.

\begin{table}[t]
           \centering
           \caption{Accuracy for the European LV feeder - without decomposition. SM error is 1\%, no impedance errors. MILP-SE-PI calculations are cut off after 2.5 hours.}
           \begin{tabular}{l||c|c|c|c}
           \hline
           \multicolumn{1}{c||}{}&\multicolumn{4}{c}{Meas. Time Length (\#time steps)} \\
           
 & 5 & 10 & 30 & 60  \\
          \hline
MILP-SE-PI & 98.9 & 99.3 & 100.0 & 100.0   \\
\hline
Correlation & 14.4 & 24.4 & 50.2 & 44.5 \\
\hline
Regression & 29.1 & 33.8 & 39.5 & 43.1 \\
\hline
\end{tabular}\label{tab:partial_results_55loads}
\end{table}

Section~\ref{sec:results_decomposed} contains complete numerical results for the decomposed European LV Feeder. Since the decomposition requires modelling three-phase (equivalent) users, the next section first presents some numerical considerations on these.

\subsection{Joint Presence of Single- and Three-phase Users}\label{sec:three_phase_results}

\begin{figure}[b]
\centering
\includegraphics[width=0.45\textwidth]{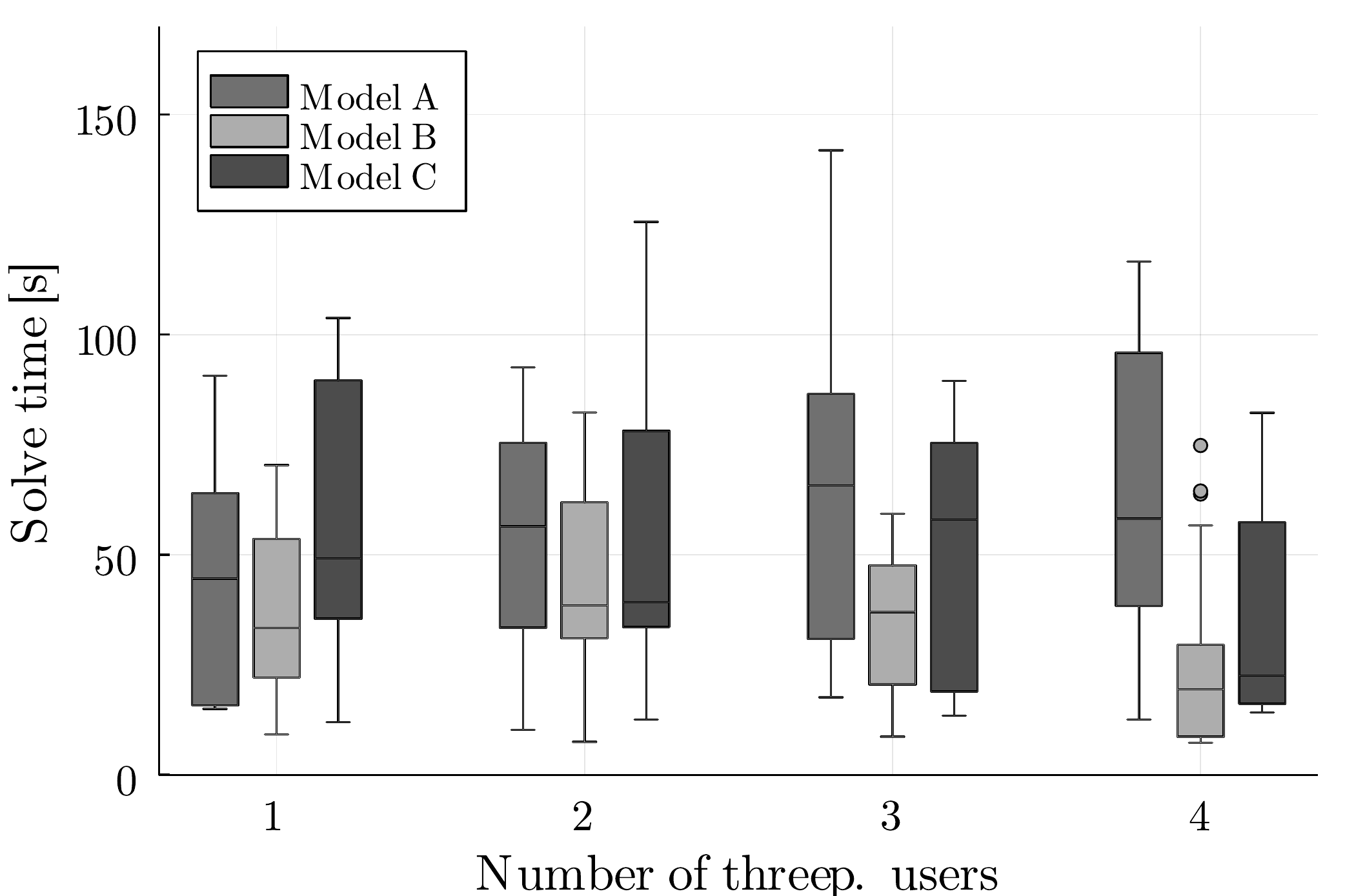}
\caption{Solver time for three-phase users analysis. Model C outliers with 1 three-phase user omitted for ease of visualization: 211, 217, 234, 343, 446, 464 s. }
\label{fig:three-phase-analysis}
\end{figure}

 \begin{table*}[b]
           \centering
           \caption{PI accuracy for the European LV Feeder (55 users), values in cells are percentages.}
           \begin{tabular}{l|l||c|c|c|c|c||c|c|c|c|c||c|c|c|c|c}
           \hline

           \multicolumn{2}{c ||}{SM Error $\rightarrow$}&\multicolumn{5}{c||}{0.5\%}&\multicolumn{5}{c||}{1\%}&\multicolumn{5}{c}{5\%} \\

           \hline

           \multicolumn{2}{c|}{}&\multicolumn{15}{c}{Meas. Time Length (\#time steps)} \\
          
           \hline
          
          \multicolumn{2}{c|}{Impedance err. $\downarrow$} & 10 & 20 & 40 & 60 & 80 & 10 & 20 & 40 & 60 & 80 & 10 & 20 & 40 & 60 & 80 \\
          \hline
          \hline
          
\multirow{2}{*}{MILP} & 0\% & 100.0 &  100.0 & 100.0 & 99.6 & 100.0 &                                 100.0 &   99.5 & 100.0 & 99.6 & 99.6 &                                 77.5 &   94.9 &  99.3 & 96.5 & 92.2 \\
\multirow{2}{*}{SE-PI}
                     & 10\% & 100.0 &  100.0 & 100.0 & 99.6 & 100.0 &   
                              100.0 &  100.0 & 100.0 & 99.6 & 100.0 &   
                               77.5 &   93.3 & 98.5 & 97.5 & 93.6 \\
                     & 20\% & 100.0 &  100.0 & 100.0 &  99.3 & 100.0 & 
                              100.0 &  100.0 & 100.0 & 99.3 & 99.6 &   78.9 & 94.2 & 99.3 & 96.5 & 95.5  \\
                    & 50\% &  100.0 & 100.0 & 100.0 & 99.6 & 99.6 & 
                              100.0 & 100.0 & 100.0 & 99.5 & 100.0 & 76.4 & 94.2 & 98.9 & 96.0 & 92.7  \\
\hline
Corr. & - & 23.3 & 37.1 & 56.9 & 57.6 & 92.0 & 24.4 & 33.8 & 42.7 & 44.5 & 70.7 & 23.5 & 28.9 & 33.1 & 36.7 & 48.2 \\
\hline
Regr. & - & 37.1 & 33.3 & 45.8 & 61.1 & 72.2 & 33.8 & 31.1 & 37.5 & 43.1 & 45.6 & 34.0 & 35.5 & 33.6 & 35.6 & 33.1 \\
\hline
\end{tabular}\label{tab:results_eulv}
\end{table*}

In the previous test cases, only single-phase users are present. This analysis is performed on Network 4, Feeder 5 of the ENWL database, which originally has 20 single-phase users. The following groups of single-phase users are aggregated into up to four three-phase users: (``8", ``17", ``10"), (``2", ``6", ``16"), and (``19", ``9", ``15"), (``13", ``11", ``3"). This is a realistic test case, as many European feeders only have limited three-phase connections with respect to the number of single-phase ones. Fig.~\ref{fig:three-phase-analysis} illustrates the computational time for the different three-phase user models described in Section~\ref{sec:mathematical_model}. The accuracy is 100\% for all models. The results refer to 6 random combinations of SM errors and the following number of timeseries lengths: 5, 10, 20. Impedance errors are not considered. It can be observed that the most performant model is, in general, model B. It is not surprising that model B performs better than model A, as constraint~\eqref{eq:model_b} implies that the former has a smaller search space. It is less obvious that model B is also better than model C, as the latter has fewer binaries. Nevertheless, it is not unprecedented that equivalent formulations with more binary variables perform better~\cite{KlotzMILP}. Here, the comparison between the methods is empirical. Deriving theoretical insights to explain these numerical differences is beyond the scope of the paper. Nevertheless, the presolver output for a 5-timestep calculation is reported in Table~\ref{tab:presolver}. It can be observed that while model B is larger than model C before presolving, the numbers of rows and nonzeros of the former reduce significantly, and become lower than for model C. Thus, it appears that something in the structure of model B facilitates the model size reduction at the presolving stage.

Nevertheless, for increasing number of three-phase users in the ``user mix", the solve time with Model C reduces. In the next section, we observe that for a network with three-phase users only, this becomes the most performant model.

\subsection{Three-phase Users Only}\label{sec:three_phase_only}

\begin{figure}[t]
\centering
\includegraphics[width=0.45\textwidth]{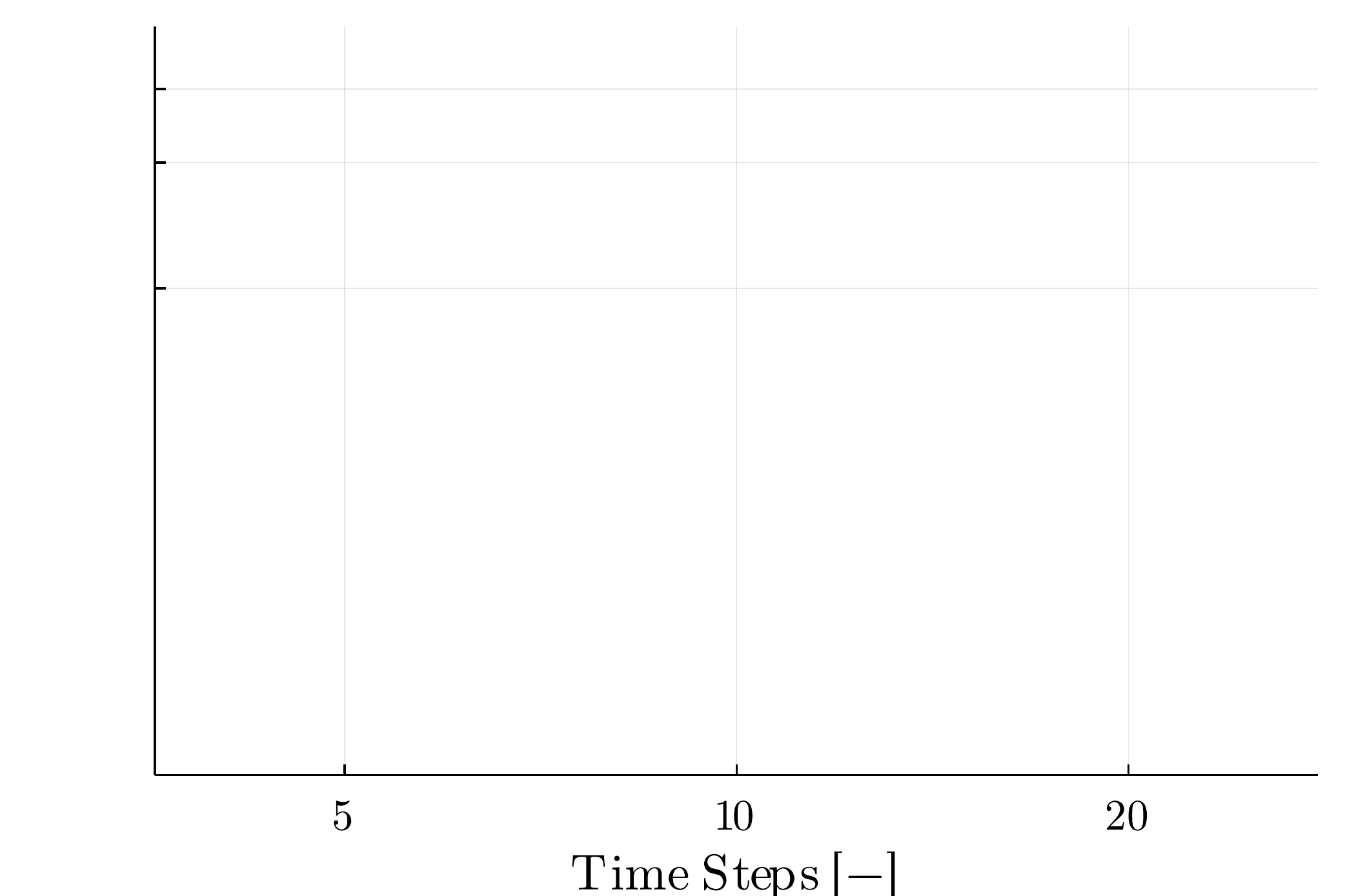}
\caption{Solve time for a six three-phase users case, logarithmic scale. }
\label{fig:threep-only}
\end{figure}

Fig.~\ref{fig:threep-only} shows the solve time for a feeder with six three-phase users. The test case is obtained by joining three copies of Feeder 2, Network 11 from the ENWL database. Each boxplot is made of 6 random combinations of SM errors. In three of the combinations with 10 time steps, Model C takes longer than 40 seconds, which increases the mean value and the spread of the boxplot. In all other cases, Model C outperforms model B (which in turn largely outperforms Model A as before, so this is left out of the present analysis).

This suggests that Model C is best employed in three-phase networks, such are some LV feeders in Northern Europe, and MV feeders in general. In these cases, PI has a different significance than in a network where single-phase users are present, and it is tantamount to ensuring that the three-phase meters are ``aligned", i.e., measurement channels are associated to the appropriate phase.

\begin{table*}[h]
\centering
\small
\caption[Presolve results for model B and C]{Presolve results for model B (fastest) and model C (fewer binaries).}
\begin{tabular}{ c c c|c|c|c } 
\hline
model & binaries/threep. usrs & & rows & colmns & nonzeros \\
\hline
\multirow{2}{*}{Model B} & \multirow{2}{*}{60/1} & before presolve  & 7193 & 4800 & 20919 \\ 
 & & after presolve & \textbf{4129} & 2346 & \textbf{13736} \\ \hline
\multirow{2}{*}{Model C} & \multirow{2}{*}{57/1} & before presolve & 7128 & 4587 & 20817  \\ 
& & after presolve & \textbf{4291} & 2261 & \textbf{14821} \\ \hline \hline
\multirow{2}{*}{Model B} & \multirow{2}{*}{60/2} & before presolve  & 7346 & 4950 & 21408 \\ 
 & & after presolve & \textbf{4043} & 2257 & \textbf{13490}  \\ \hline
\multirow{2}{*}{Model C} & \multirow{2}{*}{54/2} & before presolve & 7216 & 4524 & 21204  \\ 
& & after presolve & \textbf{4349} & 2078 & \textbf{15608} \\ \hline \hline
\multirow{2}{*}{Model B} & \multirow{2}{*}{60/3} & before presolve  & 7499  & 5100  & 21897 \\ 
 & & after presolve & \textbf{3995} & 2202 & \textbf{13366}  \\ \hline 
\multirow{2}{*}{Model C} & \multirow{2}{*}{51/3} & before presolve & 7304 & 4461 & 21591   \\ 
& & after presolve & \textbf{4407}  & 1895  & \textbf{16396}  \\ 
\hline
\hline
\multirow{2}{*}{Model B} & \multirow{2}{*}{60/4} & before presolve  &   7640 & 5250 & 22350
  \\ 
 & & after presolve & \textbf{4025} &	2256 &	\textbf{13350}
  \\ \hline 
\multirow{2}{*}{Model C} & \multirow{2}{*}{48/4} & before presolve & 7632 &	4398 & 23658
   \\ 
& & after presolve & \textbf{4587} &	1863 &	\textbf{22158}
  \\ 
\hline
\end{tabular}\label{tab:presolver}
\end{table*}

\subsection{European LV Feeder - With Decomposition}\label{sec:results_decomposed}

The European LV Test Feeder is subdivided in three parts, as shown in Fig.~\ref{fig:EULVparts}, and the equivalent three-phase users are included with model ``B", since this was the faster to solve combinations of single- and three-phase users in the previous experiments. Sub-part A has 23 users (69 binaries), sub-part B has 13 users (39 binaries), and sub-part C has 19 single-phase users (57 binaries). In practice, sub-part C is solved after the addition of the 2 three-phase equivalent users of the other sub-parts (75 binaries). The accuracy results are reported in Table~\ref{tab:results_eulv}. Overall, the accuracy is slightly lower than in the 30-user feeder case, which is reasonable as the system is bigger, more complex and with more users, but the general conclusions remain the same. Such a decrease in accuracy on the larger feeder has also been observed in~\cite{Heidari-Akhijahani2021}. As before, the accuracy of the MILP-SE-PI outperforms the other methods for the same measurement time series length, and impedance errors seem to have little impact on the accuracy, whereas SM errors have a bigger influence. 

Table~\ref{tab:results_eulv} shows results for 60 time steps that are worse than those for 40 time steps. While counter-intuitive, it is possible that in some instances the extra time steps carry errors that compromise the model rather than compensate for them with more information. A more informed choice or filtering of the measurement time steps used for the analysis could lead to more confidence on the accuracy results and is an interesting topic for future research work. With a 5\% SM error, the PI accuracy is never 100\%, as some of the 10 random instances fail to achieve 100\%. This last problem could also be solved with a better choice of measurement time steps, but, in any case, SM accuracy is usually subject to government regulation, as the readings are used for billing purposes. As such, their accuracy is typically in the class of 0.5\%~\cite{Irwin}, and the 5\% case is more illustrative than likely.

\begin{figure}[t]
\centering
\includegraphics[width=0.45\textwidth]{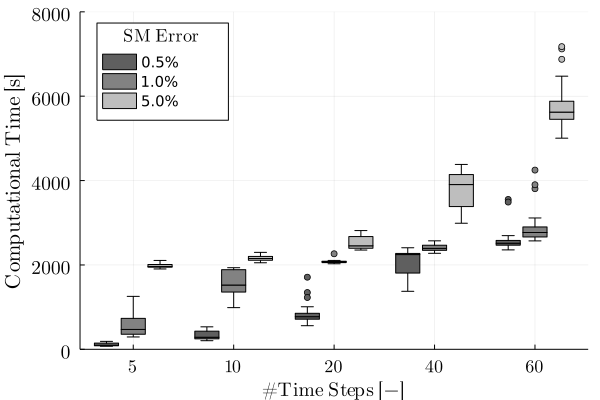}
\caption{MILP-SE-PI solver time for the EU LV Feeder.}
\label{fig:solve_time_55l}
\end{figure}

Fig.~\ref{fig:solve_time_55l} shows the computational time required to solve the European LV Feeder. This is the sum of the solve time of sub-part A, sub-part B and sub-part C with the equivalent three-phase users. While they are solved sequentially here, sub-part A and B could be solved in parallel, reducing the total time. Unsurprisingly, the solve times are longer than for the 30-user feeder. However, 24\% of the overall instances solve in less than 15 minutes, and 88\% in less than one hour, and even the slowest instance calculates again in shorter time than the extra measurement acquisition time that would be required for the other two methods to achieve the same accuracy. As with the other feeder, larger SM errors imply longer solve times.

\subsection{Summary and Notes on the MILP-SE-PI Application}

To conclude, the MILP-SE-PI is more accurate than the reference methods, in all scenarios. The scalability of MILP problems is a legitimate concern, and has been thoroughly addressed in this work. The MILP-SE-PI solves in short time, especially if combined with the network decomposition. Note that a large amount of simulations has been performed for this paper, whereas in real-life only one would be required per feeder, making time constraints less stringent. 

While the MILP-SE-PI has proved very successful in the context of this paper, the authors believe that there is no absolute best PI method. The MILP-SE-PI particularly stands out in DNs that are fully observable from a SE standpoint, for short time. This can occur where SM are widespread, but are not set to log and communicate all measurements by default, and can be activated to do so for some time, with extra costs. In the authors' experience, this or similar scenarios are realistic, at least in Europe. However, despite the increasing SM installations, partially observable systems may still be present in the near future, and other methods may be preferable to the MILP-SE-PI. A nice feature of the MILP-SE-PI is that it can also be combined with other method, e.g., fixing binary variable values where correlations are high. 

In general, as the characteristics of DNs vary highly over time and location, we believe that the access to multiple PI methods is probably necessary in the practice.

Finally, PI for three-phase users has thoroughly been addressed, which is novel. Furthermore. Model C, which is originally developed for this paper, has proved most suitable than existing methods for three-phase-user-only feeders.
\section{Conclusions}\label{sec:conclusions}
This paper presents a novel and scalable phase identification method based on state estimation and mixed-integer linear programming. The method allows to accurately identify user connectivity based on smart meter measurements only. Several ways to model three-phase users are presented and compared, which, to the best of the authors' knowledge, is an original contribution. The use of state estimation filters random noise, making the method less sensitive to measurement errors. The method is demonstrated on a set of publicly available networks and user profiles, with different smart meter and cable impedance error conditions. The method can work with different measurement setups, as long as the system is observable from a state estimation standpoint. Although only demonstrated on low voltage feeders, where phase identification is most typically performed, the method is also applicable to the medium voltage grid. The proposed method has been compared to other techniques from the literature, and has shown a higher accuracy with shorter measurement acquisition times. This has the potential to perform phase identification in near-real time, accelerating the creation of detailed network models that can benefit the energy transition. 
\section*{Acknowledgment}
Part of this work has been performed during a research visit at the Zuse Institute Berlin, and we would like to thank Prof. Koch, Dr. Tawfik, Dr. Zittel, J. Pedersen and J. Clarner for the hospitality and the shared thoughts on the present topic. The visit has been supported by the Research Foundation - Flanders (FWO), grant V414421N.

\bibliographystyle{IEEEtran}
\bibliography{IEEEabrv,Bibliography.bib}

\begin{thebibliography}{10}
\providecommand{\url}[1]{#1}
\csname url@samestyle\endcsname
\providecommand{\newblock}{\relax}
\providecommand{\bibinfo}[2]{#2}
\providecommand{\BIBentrySTDinterwordspacing}{\spaceskip=0pt\relax}
\providecommand{\BIBentryALTinterwordstretchfactor}{4}
\providecommand{\BIBentryALTinterwordspacing}{\spaceskip=\fontdimen2\font plus
\BIBentryALTinterwordstretchfactor\fontdimen3\font minus
  \fontdimen4\font\relax}
\providecommand{\BIBforeignlanguage}[2]{{%
\expandafter\ifx\csname l@#1\endcsname\relax
\typeout{** WARNING: IEEEtran.bst: No hyphenation pattern has been}%
\typeout{** loaded for the language `#1'. Using the pattern for}%
\typeout{** the default language instead.}%
\else
\language=\csname l@#1\endcsname
\fi
#2}}
\providecommand{\BIBdecl}{\relax}
\BIBdecl
\renewcommand{\BIBentryALTinterwordstretchfactor}{4}

\bibitem{Liu2019}
Y.~Liu, J.~Li, and L.~Wu, ``Coordinated optimal network reconfiguration and
  voltage regulator/der control for unbalanced distribution systems,''
  \emph{IEEE Trans. Smart Grid}, vol.~10, no.~3, pp. 2912--2922, 2019.

\bibitem{Nguyen2019}
Q.~Nguyen, H.~V. Padullaparti, K.-W. Lao, S.~Santoso, X.~Ke, and N.~Samaan,
  ``Exact optimal power dispatch in unbalanced distribution systems with high
  pv penetration,'' \emph{IEEE Trans. Power Syst.}, vol.~34, no.~1, pp.
  718--728, 2019.

\bibitem{LiuReconfiguration}
B.~Liu, K.~Meng, Z.~Y. Dong, P.~K.~C. Wong, and X.~Li, ``Load balancing in
  low-voltage distribution network via phase reconfiguration: An efficient
  sensitivity-based approach,'' \emph{IEEE Trans. Power Deliv.}, vol.~36,
  no.~4, pp. 2174--2185, 2021.

\bibitem{Gutierrez-Lagos}
L.~Gutierrez-Lagos and L.~F. Ochoa, ``Opf-based cvr operation in pv-rich
  mv–lv distribution networks,'' \emph{IEEE Trans. Power Syst.}, vol.~34,
  no.~4, pp. 2778--2789, 2019.

\bibitem{Primadianto}
A.~{Primadianto} and C.~{Lu}, ``A review on distribution system state
  estimation,'' \emph{IEEE Trans. Power Syst.}, vol.~32, no.~5, pp. 3875--3883,
  2017.

\bibitem{GethCIRED23}
F.~Geth, M.~Vanin, and D.~Van~Hertem, ``Data quality challenges in existing
  distribution network datasets,'' in \emph{Accepted for CIRED Conference
  2023}, vol. 2023, 2023.

\bibitem{CairdPatent}
K.~Caird, ``Meter phase identification,'' US Patent Application 20100164473,
  Patent No. 12/345\,702, January, 2010.

\bibitem{WenmPMUs}
M.~H. Wen, R.~Arghandeh, A.~von Meier, K.~Poolla, and V.~O. Li, ``Phase
  identification in distribution networks with micro-synchrophasors,'' in
  \emph{Proc. IEEE PES Gen. Meeting}, Denver, CO, USA, 2015, pp. 1--5.

\bibitem{Bariya}
M.~Bariya, D.~Deka, and A.~von Meier, ``Guaranteed phase \& topology
  identification in three phase distribution grids,'' \emph{IEEE Trans. Smart
  Grid}, vol.~12, no.~4, pp. 3605--3612, 2021.

\bibitem{AryaMIP}
V.~Arya, D.~Seetharam, S.~Kalyanaraman, K.~Dontas, C.~Pavlovski, S.~Hoy, and
  J.~R. Kalagnanam, ``Phase identification in smart grids,'' in \emph{Proc.
  IEEE Int. Conf. Smart Grid Commun.}, Brussels, Belgium, 2011, pp. 25--30.

\bibitem{Pappu2018}
S.~J. Pappu, N.~Bhatt, R.~Pasumarthy, and A.~Rajeswaran, ``Identifying topology
  of low voltage distribution networks based on smart meter data,'' \emph{IEEE
  Trans. Smart Grid}, vol.~9, no.~5, pp. 5113--5122, 2018.

\bibitem{Xu2018}
M.~Xu, R.~Li, and F.~Li, ``Phase identification with incomplete data,''
  \emph{IEEE Trans. Smart Grid}, vol.~9, no.~4, pp. 2777--2785, 2018.

\bibitem{Tang2018}
X.~Tang and J.~V. Milanovic, ``Phase identification of {LV} distribution
  network with smart meter data,'' in \emph{Proc. IEEE PES Gen. Meeting},
  Portland, OR, USA, 2018, pp. 1--5.

\bibitem{Jayadev}
S.~P. Jayadev, A.~Rajeswaran, N.~P. Bhatt, and R.~Pasumarthy, ``A novel
  approach for phase identification in smart grids using graph theory and
  principal component analysis,'' in \emph{Proc. American Control Conference},
  Boston, MA, USA, 2016, pp. 5026--5031.

\bibitem{GonzalezCagigal}
M.~González-Cagigal, J.~Rosendo-Macías, and A.~Gómez-Expósito,
  ``Application of nonlinear kalman filters to the identification of customer
  phase connection in distribution grids,'' \emph{International Journal of
  Electrical Power \& Energy Systems}, vol. 125, p. 106410, 2021.

\bibitem{Arya2013}
V.~Arya and R.~Mitra, ``Voltage-based clustering to identify connectivity
  relationships in distribution networks,'' in \emph{Proc. IEEE International
  Conf. on Smart Grid Comm.}, Vancouver, BC, Canada, 2013, pp. 7--12.

\bibitem{Wang2016}
W.~Wang, N.~Yu, B.~Foggo, J.~Davis, and J.~Li, ``Phase identification in
  electric power distribution systems by clustering of smart meter data,'' in
  \emph{Proc. IEEE International Conf. on Machine Learning and Applications},
  Anaheim, CA, USA, 2016, pp. 259--265.

\bibitem{Simonovska}
A.~Simonovska and L.~F. Ochoa, ``Phase grouping in pv-rich lv feeders: Smart
  meter data and unconstrained k-means,'' in \emph{Proc. IEEE PowerTech},
  Madrid, Spain, 2021, pp. 1--6.

\bibitem{Short2013}
T.~A. Short, ``Advanced metering for phase identification, transformer
  identification, and secondary modeling,'' \emph{IEEE Trans. Smart Grid},
  vol.~4, no.~2, pp. 651--658, 2013.

\bibitem{Olivier}
F.~Olivier, A.~Sutera, P.~Geurts, R.~Fonteneau, and D.~Ernst, ``Phase
  identification of smart meters by clustering voltage measurements,'' in
  \emph{Proc. Power Systems Computation Conf.}, Dublin, Ireland, 2018, pp.
  1--8.

\bibitem{PezeshkiAUPEC}
H.~Pezeshki and P.~Wolfs, ``Correlation based method for phase identification
  in a three phase {LV} distribution network,'' in \emph{Proc. 22nd
  Australasian Universities Power Engineering Conference}, Bali, Indonesia,
  2012, pp. 1--7.

\bibitem{PezeshkiISGT}
H.~Pezeshki and P.~J. Wolfs, ``Consumer phase identification in a three phase
  unbalanced {LV} distribution network,'' in \emph{Proc. IEEE PES Innovative
  Smart Grid Technologies Europe}, Berlin, Germany, 2012, pp. 1--7.

\bibitem{Foggo2020}
B.~Foggo and N.~Yu, ``Improving supervised phase identification through the
  theory of information losses,'' \emph{IEEE Trans. Smart Grid}, vol.~11,
  no.~3, pp. 2337--2346, 2020.

\bibitem{Therrien}
F.~Therrien, L.~Blakely, and M.~J. Reno, ``Assessment of measurement-based
  phase identification methods,'' \emph{IEEE Open Access Journal of Power and
  Energy}, vol.~8, pp. 128--137, 2021.

\bibitem{Hoogsteyn2022}
A.~Hoogsteyn, M.~Vanin, A.~Koirala, and D.~Van~Hertem, ``Low voltage customer
  phase identification methods based on smart meter data,'' in \emph{Accepted
  for Power Syst. Comput. Conf.}, 2022, pp. 1--7.

\bibitem{Liao2019}
Y.~Liao, Y.~Weng, G.~Liu, Z.~Zhao, C.-W. Tan, and R.~Rajagopal, ``Unbalanced
  multi-phase distribution grid topology estimation and bus phase
  identification,'' \emph{IET Smart Grid}, vol.~2, no.~4, pp. 557--570, 2019.

\bibitem{Deka2020}
D.~Deka, M.~Chertkov, and S.~Backhaus, ``Topology estimation using graphical
  models in multi-phase power distribution grids,'' \emph{IEEE Trans. Power
  Syst.}, vol.~35, no.~3, pp. 1663--1673, 2020.

\bibitem{Fernandes}
T.~Ramos~Fernandes, B.~Venkatesh, and M.~Cortes~de Almeida, ``Distribution
  system topology identification via efficient milp-based wlav state
  estimation,'' \emph{IEEE Trans. Power Syst. (Early Access)}, pp. 1--1, 2022.

\bibitem{Wang2020}
W.~Wang and N.~Yu, ``Maximum marginal likelihood estimation of phase
  connections in power distribution systems,'' \emph{IEEE Trans. Power Syst.},
  vol.~35, no.~5, pp. 3906--3917, 2020.

\bibitem{Heidari-Akhijahani2021}
A.~Heidari-Akhijahani, A.~Safdarian, and F.~Aminifar, ``Phase identification of
  single-phase customers and pv panels via smart meter data,'' \emph{IEEE
  Trans. Smart Grid}, vol.~12, no.~5, pp. 4543--4552, 2021.

\bibitem{Sankur}
M.~{Sankur}, R.~{Dobbe}, E.~{Stewart}, D.~{Callaway}, and D.~{Arnold}, ``A
  linearized power flow model for optimization in unbalanced distribution
  systems,'' \emph{[math.OC],
  Available:\url{https://arxiv.org/abs/1606.04492}}, 2016.

\bibitem{Heidari-Akhijahani2019}
A.~Heidari-Akhijahani, S.~Hojjatinejad, and A.~Safdarian, ``A milp model for
  phase identification in lv distribution feeders using smart meters data,'' in
  \emph{Proc. Smart Grid Conference}, Tehran, Iran, 2019, pp. 1--6.

\bibitem{gharebaghi2019}
S.~Gharebaghi, A.~Safdarian, and M.~Lehtonen, ``A linear model for ac power
  flow analysis in distribution networks,'' \emph{IEEE Systems Journal},
  vol.~13, no.~4, pp. 4303--4312, 2019.

\bibitem{Vanin2020}
M.~Vanin, H.~Ergun, R.~D'hulst, and D.~Van~Hertem, ``Comparison of {Linear} and
  {Conic} {Power} {Flow} {Formulations} for {Unbalanced} {Low} {Voltage}
  {Network} {Optimization},'' \emph{Electric Power Systems Research}, vol. 189,
  p. 106699, Dec. 2020.

\bibitem{Vanin2021}
M.~Vanin, T.~Van~Acker, R.~D’hulst, and D.~V. Hertem, ``A framework for
  constrained static state estimation in unbalanced distribution networks,''
  \emph{IEEE Trans. Power Syst.}, vol.~37, no.~3, pp. 2075--2085, 2022.

\bibitem{JuMP}
I.~Dunning, J.~Huchette, and M.~Lubin, ``Ju{MP}: A modeling language for
  mathematical optimization,'' \emph{SIAM Review}, vol.~59, no.~2, pp.
  295--320, 2017.

\bibitem{PMD_PSCC}
D.~M. Fobes, S.~Claeys, F.~Geth, and C.~Coffrin, ``Powermodelsdistribution.jl:
  An open-source framework for exploring distribution power flow
  formulations,'' \emph{Electr. Power Syst. Res.}, vol. 189, p. 106664, 2020.

\bibitem{KlotzMILP}
E.~Klotz and A.~M. Newman, ``Practical guidelines for solving difficult mixed
  integer linear programs,'' \emph{Surveys in Operations Research and
  Management Science}, vol.~18, no.~1, pp. 18--32, 2013.

\bibitem{pglib}
\BIBentryALTinterwordspacing
S.~{Babaeinejadsarookolaee, et al.}, ``The power grid library for benchmarking
  ac optimal power flow algorithms.'' [Online]. Available:
  \url{arXiv:1908.02788v2}
\BIBentrySTDinterwordspacing

\bibitem{ENWL}
A.~{Navarro} and L.~{Ochoa}, ``Dissemination document "low voltage networks
  models and low carbon technology profiles",'' \emph{[Online].
  Available:\url{https://www.researchgate.net/publication/283569482_Dissemination_Document_Low_Voltage_Networks_Models_and_Low_Carbon_Technology_Profiles}},
  2015.

\bibitem{Claeys2021}
S.~Claeys, F.~Geth, and G.~Deconinck, ``Line parameter estimation in
  multi-phase distribution networks without voltage angle measurements,'' in
  \emph{CIRED 2021 - The 26th International Conference and Exhibition on
  Electricity Distribution}, vol. 2021, 2021, pp. 1186--1190.

\bibitem{Raggi2022}
\BIBentryALTinterwordspacing
L.~M.~R. Raggi, V.~C. Cunha, F.~C.~L. Trindade, and W.~Freitas, \emph{Smart
  Metering in Distribution Systems: Evolution and Applications}.\hskip 1em plus
  0.5em minus 0.4em\relax Cham: Springer International Publishing, 2022, pp.
  287--317. [Online]. Available:
  \url{https://doi.org/10.1007/978-3-030-90812-6_11}
\BIBentrySTDinterwordspacing

\bibitem{Gurobi}
\BIBentryALTinterwordspacing
``Gurobi optimizer reference manual v9.1.'' [Online]. Available:
  \url{https://www.gurobi.com/documentation/9.1/refman/index.html}
\BIBentrySTDinterwordspacing

\bibitem{ipopt}
A.~{W\"{a}chter} and L.~{Biegler}, ``On the implementation of an interior-point
  filter line-search algorithm for large-scale nonlinear programming,''
  \emph{Math. Program.}, vol. 106, no.~1, pp. 25--57, 2006.

\bibitem{bible}
A.~Abur and A.~G\'{o}mez-Exp\'{o}sito, \emph{Power System State Estimation:
  Theory and Implementation}, Book, CRC Press, 2004.

\bibitem{VaninImpedance}
M.~{Vanin}, F.~{Geth}, R.~{D'hulst}, and D.~{Van Hertem}, ``Combined unbalanced
  distribution system state and line impedance matrix estimation,''
  \emph{[eess.SY], Available:\url{arXiv:2209.10938}}, 2022.

\bibitem{Irwin}
L.~A. {Irwin}, ``A high accuracy standard for electricity meters,'' in
  \emph{Proc. IEEE PES Transmission and Distribution Conference and
  Exposition}, New Orleans, LA, USA, 2010, pp. 1--3.

\end{thebibliography}

\vspace{0.2cm}
\footnotesize
\noindent
\textbf{Marta Vanin} (S'19 - M'22) received her Ph.D. in Engineering Science from the KU Leuven, Belgium, in 2022, and a joint M.Sc. degree in Energy Engineering from the University of Trento and the Free University of Bolzano in 2018. She is now working as postdoctoral researcher at KU Leuven. Her research interests include power systems modelling and optimization, and distribution system state and parameter estimation in particular.\\
\vspace{0.2cm}\\
\textbf{Tom Van Acker}, born on 28.06.1990 in Roeselare, Belgium, received the M.Eng., M.Sc. and Ph.D. degrees in Electrical Engineering from KU Leuven, Leuven, Belgium in 2012, 2014 and 2020, respectively. In 2020, he was a post-doctoral researcher at KU Leuven. Since 2021, he is a power system expert at BASF Antwerp. His main areas of research interest are stochastic processes, optimization and uncertainty quantification applied in a power system context, specifically in the topics: availability, harmonics, and state estimation.  \\
\vspace{0.2cm}\\
\textbf{Reinhilde D'hulst} received a Master’s degree in Electrical Engineering in 2004, and obtained her PhD in the field of Power Electronics in 2009, both from KU Leuven, Belgium. Since 2009 she is with the Energy Technology department of the Flemish Institute for Technological Research (VITO) and EnergyVille, Belgium.  She is involved in several national as well as European research projects related to Smart Grids, a.o. Linear, EvolvDSO, SmartNet, EUSysflex. The main focus of her work is the development of smart grid solutions for electricity (distribution) grid-related issues, distribution grid modelling, simulation and optimisation, and flexibility assessment and control algorithms for demand response. \\
\vspace{0.2cm}\\
\textbf{Dirk Van Hertem} (S’02-SM’09) graduated as a M.Eng. in 2001 from the KHK, Geel, Belgium and as a M.Sc. in Electrical Engineering from KU Leuven, Belgium in 2003. In 2009, he has obtained his PhD, also from KU Leuven. In 2010, Dirk Van Hertem was a member of EPS group at the Royal Institute of Technology (KTH), in Stockholm. Since spring 2011 he is back at the University of Leuven where he is an associate professor in the ELECTA group. His special fields of interest are decision support for grid operators, power system operation and control in systems with FACTS and HVDC and building the transmission system of the future, including offshore grids and the supergrid concept.

\end{document}